\documentclass[12pt]{iopart}
\usepackage{iopams}
\usepackage[pdftex]{graphicx}
\usepackage{float}
\newcommand{\pa}{\partial}
\newcommand{\al}{\alpha}
\newcommand{\del}{\delta}

\begin{document}
\title[Wave Optics in Black Hole Spacetimes]{Wave Optics in Black Hole Spacetimes: Schwarzschild Case}
\author{Yasusada Nambu$^1$ and Sousuke Noda$^2$}
\ead{$^1$ nambu@gravity.phys.nagoya-u.ac.jp}
\ead{$^2$ noda@gravity.phys.nagoya-u.ac.jp}
\address{Department of Physics, Graduate School of Science, Nagoya 
University, Chikusa, Nagoya 464-8602, Japan}
%%

%%%
\date{July 15, 2015} % 2nd revised version
%\date{February 19, 2015} % ver. 0.95 CQG submitted version % 
%\date{February 6, 2015, ver. 0.85} % 
%\date{February 5, 2015, ver. 0.8} % 
%\date{February 4, 2015, ver. 0.7} % 
%\date{January 30, 2015, ver. 0.6} % 
%\date{January 28, 2015, ver. 0.5} % 
%\date{January 26, 2015, ver. 0.5} % 
%\date{January 23, 2015, ver. 0.4} % 
%\date{January 21, 2015, ver. 0.3} % 
%\date{January 20, 2015, ver. 0.2} % 
%\date{January 15, 2015, ver. 0.1} % 

%%%
\begin{abstract}
  We investigate the wave optics in the Schwarzschild
  spacetime. Applying the standard formalism of wave scattering
  problems, the Green function represented by the sum over the partial
  waves is evaluated using the Poisson sum formula.  The effect of
  orbiting scattering due to the unstable circular orbit for null rays
  is taken into account as the contribution of the Regge poles of the
  scattering matrix and the asymptotic form of the scattering wave is
  obtained in the eikonal limit. Using this wave function, images of
  the black hole illuminated by a point source are reconstructed. We
  also discuss the wave effect in the frequency domain caused by the
  interference between the direct rays and the winding rays.
\end{abstract}
%%%%%%%%
\submitto{\CQG}

\maketitle

%\bibliographystyle{mybib}
%%%%%%%%%%%%%%%%%%%%%%%%%%%%%%%%%%%%%%%%%%%%
\section{Introduction}

Physics of wave propagation and wave excitation in black hole
spacetimes has been investigated as a tool to comprehend properties of
black hole spacetimes. Especially, the quasi-normal mode of black
holes is directly related to the structure of black hole spacetimes
and is expected to be confirmed observationally in near future using
gravitational waves. In the high frequency eikonal limit, the
quasi-normal mode is connected to the wave scattering in the vicinity
of the peak of the effective potential and thus related to the
existence of the unstable circular orbit for null rays around the
black hole.  The geometric interpretation of the quasi-normal
frequency is as follows; its real part corresponds to the period for a
massless particle or null rays in the unstable circular orbit and its
imaginary part represents the decay rate from the unstable circular
orbit~\cite{Mashhoon1985,Cardoso2009,Yang2012,Yang2014a}.

On the other hand, also related to the existence of the unstable
circular orbit, properties of ``shadows'' of black
holes~\cite{Fro,Falcke2000,Takahashi2004a} have been studied in detail
recently. The main motivation of this subject is direct verification
of black hole spacetimes using VLBI radio telescopes. The shadow of
the black hole is defined as the region of absorption on the
observer's screen and its rim corresponds to the unstable circular
orbit of null rays around the black hole projected on the observer's
screen. Its shape can be drawn by integrating null geodesics in the
black hole spacetime (ray tracing).  Recently, we have investigated
the imaging of black holes using waves~\cite{Nambu2013a,Kanai2013};
instead of applying the ray tracing method, the image of the black
hole is obtained using the Fourier transformation of the scattering wave
at the observer. In this method, wave effects such as interference can
be included in images of black hole shadows.

Wave optical treatment of the gravitational lensing has already done
and detectability of the interference effect is discussed for the weak
gravitational
lensing~\cite{Schneider1992,Deguchi1986,Nakamura1999,Baraldo1999}.
For ordinal astrophysical sources, we cannot expect to observe the
spatial interference fringe pattern because the coherent time of the
source is too short compared to the path difference in the gravitational
lensing system. However, even for such a case, the interference effect
can be observable in the frequency domain as the oscillation of the
power spectrum. The period of the oscillation in the power spectrum is
determined by the mass of the gravitational source which gives rise to
the gravitational lensing~\cite{Gould1992}.  For the gravitational
lensing by the black hole, a new interference effect associated with
the existence of the unstable photon orbit is expected; null rays can
go around the black hole arbitrary number of times (orbiting) and
direct rays and these winding rays can interfere and the beat appears
in the power spectrum. We expect that the period of the beat is
related to the mass of the black hole and the radius of the unstable
circular orbit.

In this paper, we consider wave optics in the Schwarzschild spacetime
using a massless scalar field to investigate wave effects peculiar to
black hole spacetimes. The massless scalar field is adopted as the
benchmark treatment for the wave scattering problem by black holes and
we do not consider polarization degrees of freedom which is necessary
for the electro magnetic waves and gravitational waves.  We follow the
standard treatment of the wave scattering problem by black
holes~\cite{Andersson2000,Glampedakis2001,Decanini2011b} and derive
the asymptotic form of the scattered wave at sufficiently far from the
black hole in the high frequency eikonal limit. The wave is
represented by the sum over the partial waves which include the WKB
phase shift. Comparing to the standard treatment of the wave
scattering theory, our formulation retains the next to leading order
contribution in the phase of the scattering wave at large $r$, and
this refinement makes the sum over partial waves converge even for the
scattering caused by the long range $1/r$ potential.

We shows that the gravitational lens equation in the black hole
spacetime can be derived by taking the short wavelength limit of the
scattering wave.  This lens equation is an extension of the relation
between the scattering angle and the derivative of the WKB phase shift
which is derived by the stationary phase
method~\cite{Ford1959a,Berry1969}.  To evaluate the sum over partial
waves, we apply the complex angular momentum (CAM)
method~\cite{Newton1982,Andersson1994a,Decanini2003,Decanini2010,Decanini2011b}.  The scattering
wave is decomposed to the direct part and the winding part. The
winding part can be evaluated using Regge poles which are associated
with the existence of the unstable circular orbit around the black
hole.
 
The outline of this paper is as follows. In Sec.~II, we introduce the
WKB Green function and derive the lens equation from the Green
function by the stationary phase method. In Sec.~III, we shortly
review the S-matrix and the Regge poles. We evaluate the sum over
partial waves and obtain the asymptotic form of the scattering wave in
Sec.~IV. As  applications, we consider images of the black hole and
the interference effect in the power spectrum in Sec.~V. Sec.~VI is
devoted to summary.  We use units in which $c=\hbar=G=1$ throughout
this paper.

%%%%%%%%%%%%%%%%%%%%%%%%%%%%%%%%%%%%%%%%%%%%
\section{Wave scattering by a black hole}
We investigate the wave scattering problem by the black hole in
the high frequency eikonal limit. The high frequency condition is
$M\omega\gg 1$ where $M$ is the mass of a black hole. Thus this
condition states that the wave length is sufficiently shorter than the
 size of the black hole. The eikonal limit is the angular quantum
 number $\ell$ of the wave is larger than unity $\ell\gg 1$. Under
 these conditions, we follow the standard approach of
the wave scattering theory; decompose scattered waves to the sum over
partial waves and introduce the phase shift which represents the
properties of the scatterer.

%%%%%%%%%%%%%%%%%%%%%%%%
\subsection{Wave equation and the WKB phase shift}
Let us consider the scalar wave equation $\square\Phi=0$ in the
Schwarzschild spacetime. The metric is
%%%
\begin{equation}
  ds^2=-\left(1-\frac{2M}{r}\right)dt^2+\left(1-\frac{2M}{r}\right)^{-1}dr^2
  +r^2d\Omega^2.
\end{equation}
%%%%
As the geometry has the spherical symmetry, for the wave configuration
with axial symmetry about the $z$-axis, the scalar field $\Phi$ is
separable as
%%%
\begin{equation}
   \Phi=e^{-i\omega t}R_{\ell}(r)P_{\ell}(\cos\theta).
\end{equation}
%%%
The radial wave function $R_\ell$ and the angular wave function
$P_\ell$ satisfy
%%%
\begin{eqnarray}
    &\frac{d}{dr}\left(\Delta\frac{dR_\ell}{dr}\right)
    +\left(\frac{r^4\omega^2}{\Delta}-\ell(\ell+1)\right)R_\ell=0,\\
    &\frac{1}{\sin\theta}\frac{d}{d\theta}
    \left(\sin\theta\frac{dP_\ell}{d\theta}\right)+\ell(\ell+1)P_\ell=0,
\end{eqnarray}
%%%
where $\Delta=r^2-2Mr$. We consider the WKB solution of the radial
equation with the assumption of the high frequency eikonal limit
$M\omega\gg 1$, $\ell\gg 1$.  By introducing a new radial function
$\psi_\ell=r R_\ell$ and the tortoise coordinate $r_*=\int
dr\frac{r^2}{\Delta}=r+2M\ln(r/(2M)-1)$, the radial equation becomes
%%%
\begin{eqnarray}
 &\frac{d^2\psi_\ell}{dr_*^2}+Q\psi_\ell=0, \label{eq:req2}\\
 &Q=\omega^2-\left(\frac{\ell(\ell+1)}{r^2}+\frac{2M}{r^3}\right)
 \left(1-\frac{2M}{r}\right). \nonumber
\end{eqnarray}
%%%
The lowest order WKB solution of this equation is
%%%
\begin{equation}
  \psi_\ell=Q^{-1/4}\exp\left[i\int^{r_*} dr_*\,Q^{1/2}\right]\propto e^{iS_r}.
\end{equation}
%%%
For the angular wave function, which is the Legendre function in the
present case, it also can be written in the WKB form $P_\ell\propto
e^{iS_\theta}$. For $\ell, M\omega\gg 1$, the phase functions $S_r$
and $S_\theta$ satisfy the following Hamilton-Jacobi equations
%%%
\begin{eqnarray}
    &\left(\frac{dS_r}{dr}\right)^2-\frac{1}{\Delta^2}
    \left(r^4\omega^2-\ell(\ell+1)\Delta\right)=0,\\
    &\left(\frac{dS_\theta}{d\theta}\right)^2-\ell(\ell+1)=0.
\end{eqnarray}
%%% 
These equations are the same as the Hamilton-Jacobi equations for null
rays in the Schwarzschild spacetime.  The solutions of
these equations are
%%%
\begin{equation}
 S_r(r)=\int^r dr\frac{\sqrt{\mathcal{R}}}{\Delta},\quad
 S_\theta(\theta)=\int^\theta d\theta\,L=L\,\theta,
\end{equation}
%%%
where $\mathcal{R}(r)=\omega^2r^4-L^2\Delta$ and we have replaced
$\ell(\ell+1)$ by $L^2\equiv(\ell+1/2)^2$, which improves the accuracy
of the WKB approximation; this replacement ensures the phase shift
introduced in (\ref{eq:phaseshift}) zero for the flat limit $M=0$.
Following the standard Hamilton-Jacobi theory, by differentiating the
function $S_r+S_\theta$ with respect to the angular momentum $L$, the
equation for the trajectory of null rays is derived as
%%%
\begin{equation}
 -L\int_{r_i}^{r_f}\frac{dr}{\sqrt{\mathcal{R}}}=-(\theta_f-\theta_i).
\end{equation}
%%%
From this equation, the deflection angle for null rays is obtained by
taking $r_{i},r_{f}\rightarrow\infty$:
%%%
\begin{equation}
 \label{eq:dfangle}
 \Theta=\pi-(\theta_f-\theta_i)=
 \pi-2L\int_{r_0}^\infty\frac{dr}{\sqrt{\mathcal{R}}},
\end{equation}
%%%
where $r_0$ is the radial turning point determined by
$\mathcal{R}(r_0)=0$.  

We rewrite the WKB radial function using the phase shift. For this
purpose, we consider the solution of the radial equation
(\ref{eq:req2}) with the following boundary condition:
%%%
\begin{equation}
  \label{eq:uin}
  u_{\mathrm{in}}\sim
  \cases{
     e^{-i\omega r_*}, &$ r_*\rightarrow-\infty$,\\
     A_{\mathrm{out}}\,e^{i\omega r_*}+A_{\mathrm{in}}\,e^{-i\omega r_*},
     &$ r_*\rightarrow+\infty$.\\
  }
\end{equation}
%%%%
The phase shift is defined by
%%%
\begin{equation}
  e^{2i\del_\ell}=-(-)^{\ell}\frac{A_{\mathrm{out}}}{A_\mathrm{in}}.
\end{equation}
%%%
 Then the wave at
$r\rightarrow+\infty$ is expressed as
%%%
\begin{eqnarray}
  u_{\mathrm{in}}&\sim A_{\mathrm{out}}\,e^{i\omega
    r_*}+A_{\mathrm{in}}\,e^{-i\omega r_*}  \nonumber\\
  &\propto \sin\left(\omega r_*+\del_\ell-\frac{\pi\ell}{2}\right).
\end{eqnarray}
%%%
For $\del_\ell=0$, $u_{\mathrm{in}}$ corresponds to the spherical wave
in the flat spacetime $j_\ell(\omega r)\sim\sin(\omega r-\pi\ell/2)$.
Within the WKB approximation, the analytic
continuation around the turning point yields the radial function for large
$r$ as
%%%
\begin{eqnarray}
  u_{\mathrm{in}}(r)&\approx\sin\left[\int_{r_\mathrm{t}}^{r_*}
    dr_*\,Q^{1/2}+\frac{\pi}{4}\right] \nonumber \\
  &=\sin\left[\omega
    r_*+\del_\ell-\frac{\pi\ell}{2}-
    \int_r^\infty dr\frac{r^2}{\Delta}\left(Q^{1/2}-\omega\right)\right],
    \label{eq:waveR}
\end{eqnarray}
%%%
where $r_\mathrm{t}$ is the turning point determined by
$Q(r_\mathrm{t})=0$.  Hence the WKB phase shift in the Schwarzschild
spacetime is obtained as~\cite{Glampedakis2001}
%%%
\begin{equation}
\label{eq:phaseshift}
\del_\ell=
 \int_{r_\mathrm{t}}^\infty
 dr\frac{r^2}{\Delta}\left(Q^{1/2}-\omega\right)
 -\omega r_{\mathrm{t}*}+\frac{\pi}{2}\left(\ell+\frac{1}{2}\right).
\end{equation}
%%%
For small $\ell$ less than a some critical value $\ell_c$, there
exists no real solution for $Q(r)=0$ and (\ref{eq:phaseshift}) has no
meaning. However, by performing a suitable analytic extension of the
formula, it can be shown that the phase shift acquires an imaginary
part which represents the absorption of waves by the black hole.  In
the eikonal limit $\ell\gg 1$, $Q(r)\approx \mathcal{R}(r)/r^4$ and
the phase shift is
%%%
\begin{equation}
 \del_\ell=\int_{r_0}^\infty
 dr\left(\frac{\sqrt{\mathcal{R}}}{\Delta}-\frac{r^2}{\Delta}\omega\right)
 -\omega r_{0*}+\frac{\pi}{2}\left(\ell+\frac{1}{2}\right),
\end{equation}
%%%
and by differentiating with respect to the angular momentum $\ell$,
%%%
\begin{equation}
 \frac{d\del_\ell}{d
   \ell}\approx \frac{\pi}{2}-\left(\ell+\frac{1}{2}\right)
 \int_{r_0}^\infty\frac{dr}{\sqrt{\mathcal{R}}}.
\end{equation}
%%%
Comparing to Eq.~(\ref{eq:dfangle}), we thus obtain the well known
relation between the classical deflection angle and the WKB phase
shift~\cite{Newton1982}
%%%
\begin{equation}
 \label{eq:deflection}
 \Theta=2\frac{d\del_\ell}{d\ell}.
\end{equation}
%%%
For large $r$, the integral in the phase of the radial wave function
(\ref{eq:waveR}) can be approximated to be
%%%
\begin{equation}
-\int_{r}^\infty
dr\left(1+\frac{2M}{r}\right)
\left(Q^{1/2}-\omega\right)\approx
 \int_r^\infty dr \frac{(\ell+1/2)^2}{2\omega r^2}=
 -\frac{(\ell+1/2)^2}{2\omega r},
\end{equation}
%%%%
where we assume that the function $Q^{1/2}$ is real so $\ell$ must
satisfies $\sqrt{2}\,\omega r\ge\ell+1/2$ to make this approximation
have meaning. This imposes the maximal upper limit
$\ell_\mathrm{max}\approx \sqrt{2}\,\omega r$ for a fixed value of
$\omega r$ when we take the partial wave sum.

In the standard formulation of the wave scattering problem, this term
is treated as zero because it vanishes as $r\rightarrow \infty$ and
does not contribute to the scattering amplitude and the differential
cross section. However, as we will see, this term is necessary to
reproduce the lens equation from the wave function in the eikonal
limit and also necessary for convergence of the sum over partial
waves.

After all, for sufficiently large $r$, we obtain the following form of
the radial function in the eikonal limit
%%%
\begin{equation}
  u_{\mathrm{in}}(r)\propto \sin\left[\omega
    r_*+\frac{(\ell+1/2)^2}{2\omega
      r}+\del_\ell-\frac{\pi\ell}{2}\right]\quad\mathrm{with}\quad
  \ell\le \sqrt{2}\,\omega r.
\end{equation}
%%%
%%%%%%%%%%%%%%%%%%%%%%%%%%%%%%%%%%%%%%%%%%%%%%
\subsection{Green function}
We aim to obtain the scattered wave by a black hole; the scalar wave is
emitted from a point source and scattered by the Schwarzschild black
hole (see Fig.~1). For this purpose, let us consider the Green function for the
scalar wave. 
%%%
\begin{figure}[H]
  \centering
  \includegraphics[width=0.5\linewidth,clip]{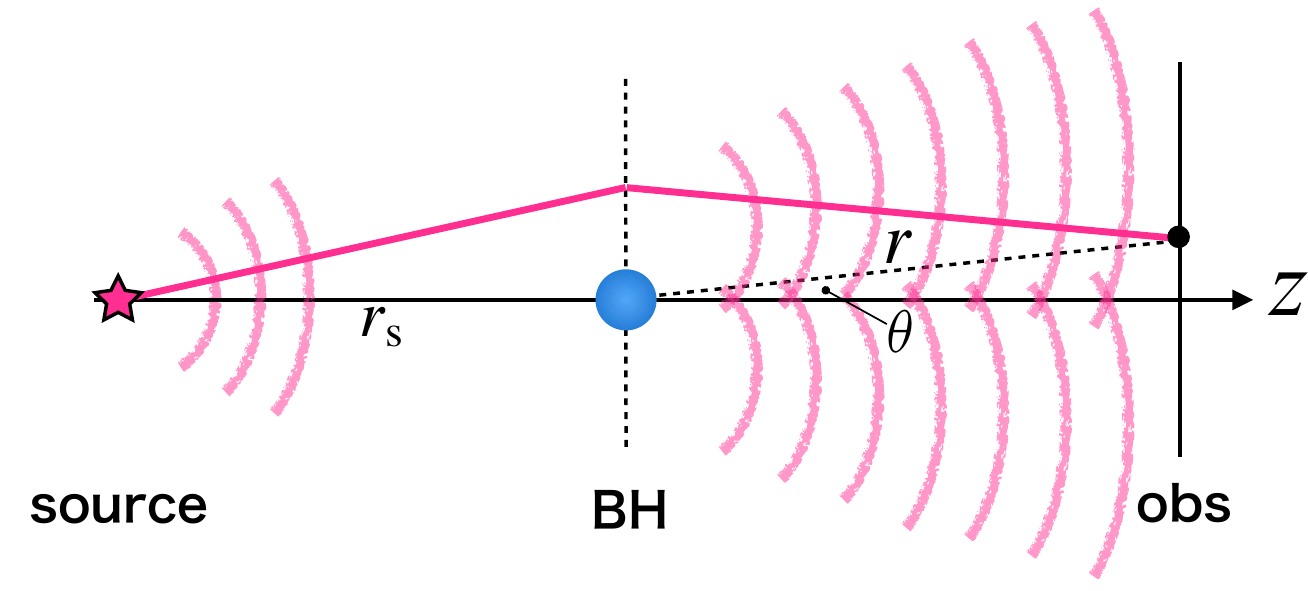}
   \caption{Configuration of wave scattering by a black hole.}
\end{figure}
%%%
\noindent 
For a monochromatic stationary wave with time dependence $e^{-i\omega
  t}$, the Green function satisfies
%%%
\begin{equation}
 -g^{00}\omega^2\Phi+\frac{1}{\sqrt{-g}}\pa_j(\sqrt{-g}g^{jk}\pa_k\Phi)=-\del^3
  (x,x_s).
\end{equation}
%%%
where $\del^3(x,x_s)$ denotes the invariant delta function
$\frac{1}{\sqrt{-g}}\del^3(x-x_s)$.  By separating the wave function
as
%%%
\begin{equation}
\Phi(r,\theta;r_s,\theta_s)=\sum_{\ell=0}^\infty\frac{2\ell+1}{4\pi rr_s}
\psi_{\ell}(r,r_s)P_{\ell
  }(\cos\theta)P_{\ell}(\cos\theta_s),
\end{equation}
%%%
the radial Green function $\psi_\ell(r,r_s)$ obeys\footnote{We use the relation
  $\sum_{\ell=0}^\infty(\ell+\frac{1}{2})P_{\ell}(\cos\theta)P_{\ell}(\cos\theta_s)=\del(\cos\theta-\cos\theta_s)$.}
%%%
\begin{equation}
 \label{eq:schw-radial}
 \frac{d^2\psi_\ell}{dr_*^2}+Q(r)\psi_\ell=-\del(r_*-r_{s*}).
\end{equation}
%%%
The solution of this equation can be constructed from a pair of
linearly independent solutions of the radial equation; one is given by
$u_\mathrm{in}$ in (\ref{eq:uin}) and the other has the following
asymptotic behavior
%%%
\begin{equation}
  \label{eq:uup}
  u_{\mathrm{up}}\sim
  \cases{
     ~B_\mathrm{in}\,e^{-i\omega r_*}+B_\mathrm{out}\,e^{i\omega r_*}, 
     & $r_*\rightarrow-\infty$,\\
     ~e^{i\omega r_*},
     & $r_*\rightarrow+\infty$.
  }
\end{equation}
%%%
Using these two independent solutions $u_{\mathrm{in}}, u_{\mathrm{up}}$, the
radial Green function can be written as
%%%
\begin{equation}
  \psi_\ell(r,r_s)=-\frac{u_{\mathrm{in}}(r_{<})\,u_{\mathrm{up}}(r_{>})}{W},
\end{equation}
%%%
where $r_{>}=\mathrm{max}(r,r_s), r_{<}=\mathrm{min}(r,r_s)$ and the
Wronskian $W$ is
%%%
\begin{equation}
  W=u_{\mathrm{in}}\frac{du_{\mathrm{up}}}{dr_*}
  -u_{\mathrm{up}}\frac{du_{\mathrm{in}}}{dr_*}=2i\omega A_{\mathrm{in}}.
\end{equation}
%%%
For large $r,r_s$ with $r_s<r$, the radial Green function is
%%%
\begin{equation}
    \label{eq:psil}
  \psi_{\ell}(r,r_s)\approx \frac{i}{2\omega}e^{i\omega r_*}\left(
    e^{-i\omega r_{s*}}-(-)^\ell e^{2i\del_{\ell}}\,e^{i\omega r_{s*}}\right).
\end{equation}
%%%
Assuming that the point source of the wave is located on $-z$
axis ($\theta_s=\pi$), the full Green function for the scattering problem is 
%%%
\begin{equation}
  \Phi(r,\theta;r_s,\theta_s)
  =\frac{i\omega}{4\pi rr_s}\sum_{\ell=0}^\infty (2\ell+1)(-)^\ell
  \psi_{\ell}(r,r_s)P_\ell(\cos\theta).
 \label{eq:wave0}
\end{equation}
%%%
 For $M=0$ (free propagating wave in the flat
spacetime), $\psi_{\ell}(r,r_s)=rr_s j_\ell(\omega r_s)h^{(1)}(\omega r)$ and
this formula reduces to the Green function of the freely propagating
wave in the flat spacetime:
%%%
\begin{equation}
  \label{eq:Phi0}
  \Phi_0
  =\frac{i\omega}{4\pi}\sum_{\ell=0}^\infty(2\ell+1)(-)^\ell
  j_\ell(\omega r_s)h^{(1)}_\ell(\omega r)P_\ell(\cos\theta)
  = \frac{1}{4\pi}\frac{e^{i\omega|\mathbf{x}-\mathbf{x}_s|}}{|\mathbf{x}-\mathbf{x}_s|}.
\end{equation}
%%%
Including the term $\frac{(\ell+1/2)^2}{2\omega r}$ in the phase of
the radial function (\ref{eq:psil}), the asymptotic form of the Green
function at large $r$ is
%%%
\begin{eqnarray}
 \fl  \Phi(r,\theta) 
\approx\frac{e^{i\omega r_{*}}}{8i\pi\omega
    rr_s}
  \sum_{\ell=0}^\infty(2\ell+1)\left[
    e^{i\left(\omega
      r_{s*}+\frac{(\ell+1/2)^2}{2\omega}(\frac{1}{r}+\frac{1}{r_s})
        +2\del_\ell\right)}-(-)^\ell e^{-i\left(\omega
        r_{s*}+\frac{(\ell+1/2)^2}{2\omega}(-\frac{1}{r}+\frac{1}{r_s})\right)
    }\right]P_\ell(\cos\theta)\nonumber\\
=\Phi_0+\frac{e^{i\omega(r_{s*}+r_*)}}{4\pi
    r_s}\frac{1}{2i\omega r}\sum_{\ell=0}^\infty(2\ell+1)
  e^{i\frac{(\ell+1/2)^2}{2\omega\tilde r
      }}(e^{2i\del_\ell}-1)
    P_\ell(\cos\theta), \label{eq:waveF0}
\end{eqnarray}
%%%
where $\Phi_0\equiv\Phi|_{\delta=0}$ represents the incident spherical
wave from the point source and we defined 
%%%
\begin{equation}
\frac{1}{\tilde r}\equiv
\frac{1}{r}+\frac{1}{r_s}.
\end{equation}
%%%
 Following the standard
prescription of preparing the plane wave in the black hole
spacetime~\cite{Futterman1987}, the spherical wave from the point
source placed at sufficiently far from the black hole is obtained by
replacing $r, r_s$ in the phase factor of (\ref{eq:Phi0}) to their
tortoise coordinate $r_*, r_{s*}$:
%%%
\begin{equation}
\Phi_0\approx \frac{e^{i\omega (r_{s*}+r_*)}}{4\pi
    (r+r_s)}e^{-i\omega\tilde r
    \frac{\theta^2}{2}}\quad\mathrm{for}\quad
  \theta\ll1~\mathrm{and~ large}~ r,r_s.
\end{equation}
%%%
At this stage, we apply the Poisson sum formula\footnote{The Poisson sum
  formula is $$
  \sum_{\ell=0}^\infty
  F(\ell+1/2)=\sum_{m=-\infty}^{+\infty}(-)^m\int_0^\infty
  F(\lambda)\,e^{i2\pi m\lambda}d\lambda. $$} to  the
   sum in (\ref{eq:waveF0}) which does not contain the phase shift:
%%%
\begin{eqnarray}
  &\sum_{\ell=0}^{\infty}(2\ell+1)e^{i\frac{(\ell+1/2)^2}{2\omega\tilde
    r}}
P_\ell(\cos\theta)  \label{eq:sum0}\\
&\qquad
=2\sum_{m=-\infty}^{+\infty}\int_0^\infty
  d\lambda\,\lambda\,e^{i\frac{\lambda^2}{2\omega\tilde r
      }}
    P_{\lambda-1/2}(\cos\theta)\,e^{2im\pi(\lambda-1/2)} \nonumber
  \\
&\qquad \approx 2\int_0^\infty
  d\lambda\,\lambda\,e^{i\frac{\lambda^2}{2\omega\tilde r
      }}J_0(\lambda\theta)
    +4\sum_{m=1}^{+\infty}\int_0^\infty
  d\lambda\,\lambda\,e^{i\frac{\lambda^2}{2\omega\tilde r
      }}
    J_{0}(\lambda\theta)\,\cos(2m\pi(\lambda-1/2))  \nonumber \\
  &\qquad =2i\omega\frac{rr_s}{r+r_s}
  e^{-i\omega\tilde r\frac{\theta^2}{2}}+O((\omega \tilde r)^0),
    \quad\mathrm{for}\quad
  \theta\ll 1, \nonumber
\end{eqnarray}
%%%
where we used the asymptotic formula for the Legendre function
%%%
\begin{equation}
 P_{\lambda-1/2}(\cos\theta)\approx\left(\frac{\theta}{\sin\theta}\right)^{1/2}
 J_0(\lambda\theta)\quad\mathrm{for}\quad\lambda\gg1,\quad \theta\neq\pi,
\end{equation}
%%%
and consider the scattering with small $\theta$ (forward direction) in
the present analysis.  We neglect $O(\omega^0)$ contribution which is higher
order in the eikonal approximation.  Therefore, for large $r$, the
Green function in the form of the partial wave sum is
%%%
\begin{eqnarray}
 \fl \Phi\approx\frac{e^{i\omega (r_*+r_{s*})}}{4\pi (r+r_s)}\left[
    e^{-i\omega\tilde r\frac{\theta^2}{2}}
    -e^{-i\omega\tilde r\frac{\theta^2}{2}}+
    \frac{r+r_s}{i\omega rr_s}\sum_\ell\left(\ell+\frac{1}{2}\right)
    e^{i\frac{(\ell+1/2)^2}{2\omega\tilde r
        }}
      \,e^{2i\del_\ell}J_0((\ell+1/2)\,\theta)\right] \nonumber\\
  =
   \frac{e^{i\omega (r_*+r_{s*})}}{4\pi i\omega rr_s}
     \sum_{\ell=0}^{\infty}
     \left(\ell+\frac{1}{2}\right)e^{i\frac{(\ell+1/2)^2}{2\omega\tilde r
        }}
      \,e^{2i\del_\ell}J_0((\ell+1/2)\,\theta),\quad\mathrm{for}
      \quad\theta\ll1.\label{eq:wave1}
\end{eqnarray}
%%%
In this formula, the factor
$e^{i\frac{(\ell+1/2)^2}{2\omega\tilde r}}$
is essential to make the infinite sum over $\ell$ finite. Without this
factor, the sum does not converge; this is well known property of the partial
wave sum for long range potential which falls off as $1/r$ at large
distances. We will discuss the issue of convergence of the partial
wave sum (\ref{eq:sum0}) and (\ref{eq:wave1}) in the next subsection.

We take a look at the scattering of the plane wave in the weak
gravitational field of a point mass and check the formula
(\ref{eq:wave1}) indeed works well. In this case, the deflection angle
is given by the Einstein formula for light deflection
%%%
\begin{equation}
    \Theta(b)=-\frac{4M}{b}\approx -\frac{4M\omega}{\ell},
\end{equation}
%%%
where $b=(\ell+1/2)/\omega$ is the impact parameter for the partial
wave. The phase shift is obtained by integrating the relation
(\ref{eq:deflection})
%%%
\begin{equation}
  \label{eq:Nphase}
    \del_\ell=\frac{1}{2}\int^{\ell}d\ell\,\Theta(\ell)=-2M\omega\ln\ell
+\mathrm{const.}
\end{equation}
%%%
By taking $r_s\rightarrow\infty$, the incident spherical wave reduces
to the plane wave. Renormalizing the over all factor as $e^{i\omega
  r_{s*}}/(4\pi r_s)\rightarrow 1$ and applying the Poisson sum
formula to (\ref{eq:wave1}),
%%%
  \begin{eqnarray}
      \Phi&=\frac{e^{i\omega r_*}}{i\omega
        r}\sum_{\ell=0}^{\infty}\left(\ell+\frac{1}{2}\right) 
      e^{i\frac{(\ell+1/2)^2}{2\omega
          r}}e^{-i4M\omega\ln\ell}J_0((\ell+1/2)\theta)  \label{eq:wave2} \\
      &=\frac{e^{i\omega r_*}}{i\omega r}\int_0^\infty d\lambda\,\lambda\,
      e^{i\frac{\lambda^2}{2\omega
          r}}\,\lambda^{-i4M\omega}J_0(\lambda\theta)+O(1/(\omega r))
      \nonumber\\
      &=e^{-i2M\omega\ln(4M\omega)}\,e^{i\omega
        r(1-\theta^2/2)}\,e^{\pi
        M\omega}\Gamma(1-i2M\omega)\,{}_1F_1\left(2iM\omega, 1,
      \frac{i}{2}r\omega\theta^2\right). \nonumber
  \end{eqnarray}
%%%%
The phase factor $e^{-i2M\omega\ln(4M\omega)}$ can be absorbed to the
constant of the phase shift. Then the obtained wave function has the
following asymptotic form for $\omega r\theta^2\gg 1$,
%%%
\begin{equation}
 \fl   \Phi\approx e^{i\omega z-2i\omega
      M\ln(\omega(r-z))}+M\frac{\Gamma(1-2iM\omega)}
    {\Gamma(1+2iM\omega)}\left(\frac{\theta}{2}\right)^{-2+4iM\omega}
    \frac{e^{i\omega(r+2M\ln(2\omega r))}}{r},
\end{equation}
%%%
where $z\approx r(1-\theta^2/2), r-z\approx r\theta^2/2$. The first
term is the incident plane wave and the second term is the scattering
wave. Thus (\ref{eq:wave2}) reproduces the scattering wave function
for $1/r$ potential (Coulomb scattering). For $M\omega\gg 1,
r\omega\theta^2=\mathrm{finite}$, corresponding to the scattering toward
the forward direction $\theta\sim0$,
%%%
\begin{equation}
    \Phi\approx e^{i\omega r(1-\theta^2/4)}\,e^{\pi
      M\omega}\,\Gamma(1-i2M\omega)
    \,J_0(\sqrt{4Mr}\,\omega\theta).
\end{equation}
%%%

%%%%%%%%%%%%%%%%%%%%
\subsection{Convergence of partial wave sum}

Here we comment on convergence of the partial wave sum
(\ref{eq:wave1}). We truncate the infinite sum up to
$\ell_\mathrm{max}$ and the phase shift is evaluated by the WKB
formula (\ref{eq:phaseshift}). As we have mentioned in the last paragraph in
Sec.~2.1, the upper bound of the partial wave sum in the eikonal
approximation must be $\ell_\mathrm{max}=\sqrt{2}\,\omega r$, and this
value becomes infinity in the eikonal limit. To check the convergence
of this sum, we first consider (\ref{eq:sum0}) which is $\del_\ell=0$ case
of (\ref{eq:wave1}). Let us consider the following quantity:
%%%
\begin{equation}
  I(\theta,\ell_\mathrm{max})=\frac{1}{2\omega
    r}\sum_{\ell=0}^{\ell_\mathrm{max}}
  (2\ell+1)\,e^{i\frac{(\ell+1/2)^2}{2\omega r}}J_0((\ell+1/2)\theta).
    \label{eq:sum1}
\end{equation}
%%%
The relation (\ref{eq:sum0}) implies this sum becomes
%%%
\begin{equation}
  |I(\theta)|=1+O((\omega r)^{-1})
\end{equation}
%%%
in the eikonal limit $\omega r\gg 1$. We present converging behavior
of this sum for $\theta=0$ case (Fig.~2).
%%%
\begin{figure}[H]
  \centering
  \includegraphics[width=0.7\linewidth,clip]{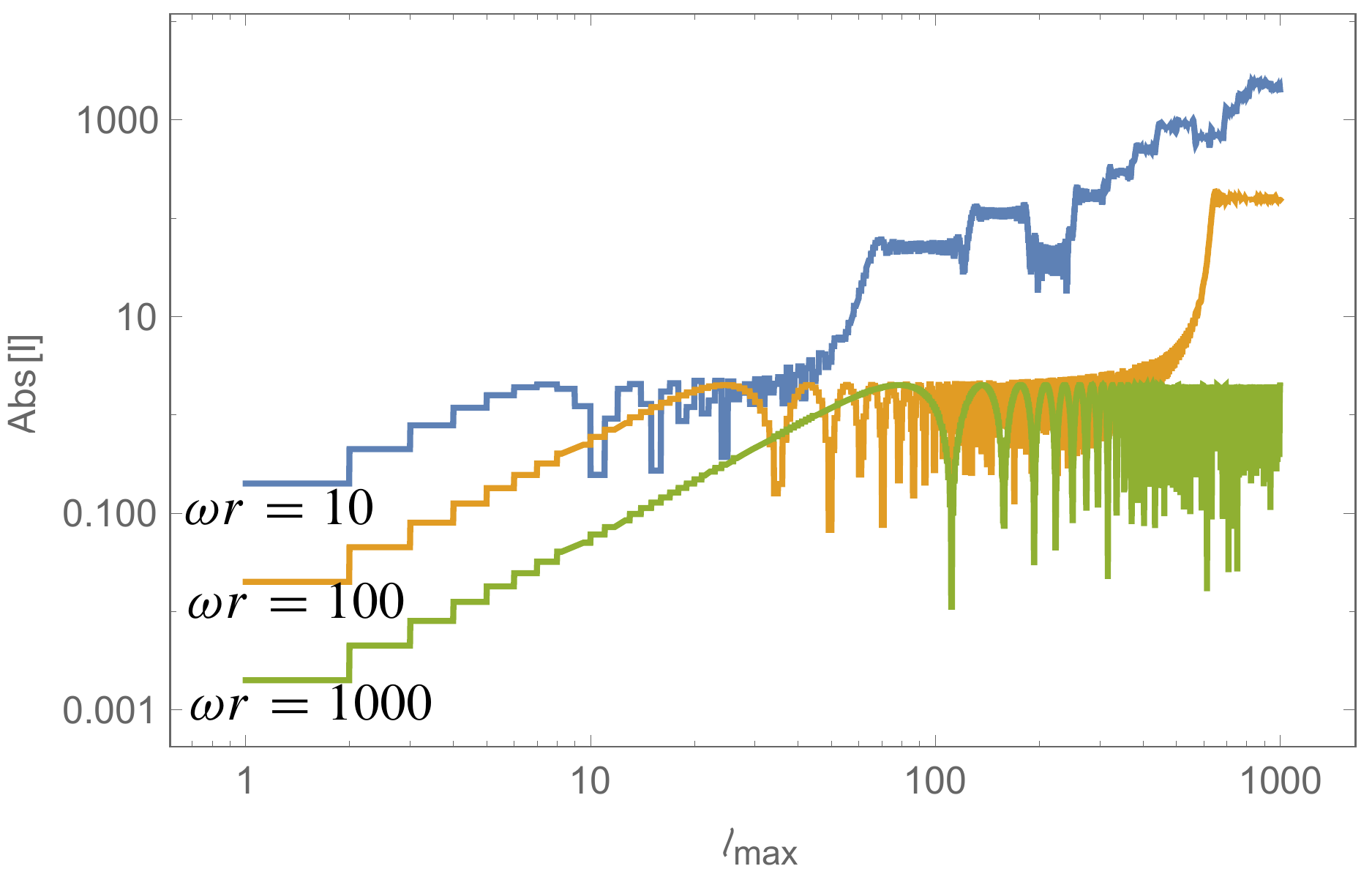}
  \caption{The partial wave sum (\ref{eq:sum0}) as a function of
    $\ell_\mathrm{max}$ for $\omega r=10,100,1000$.}
\end{figure}
%%%
\noindent
For $\ell_\mathrm{max}\approx\sqrt{2}\,\omega r$, we can observe the
absolute value of the sum oscillates around the true value predicted
by the formula (\ref{eq:sum0}). This oscillation is due to
interference between $O((\omega r)^0)$ and $O((\omega r)^{-1})$ terms
and the amplitude of the oscillation is expected to be reduced in
$\omega r\gg 1$ limit. Thus we expect that the partial wave sum with
$\ell_\mathrm{max}=\sqrt{2}\,\omega r$ converges in the eikonal limit.

We then checked numerically the convergence of the sum of
(\ref{eq:wave1}) for the weak gravitational field case that is
equivalent to the Coulomb scattering.  We truncate the infinite sum up to
the value $\ell_\mathrm{max}$ and the phase shift is evaluated by the
formula (\ref{eq:Nphase}). The convergence of the sum (\ref{eq:wave2})
for $\theta=0 $ is shown in Fig.~3:
%%%
\begin{figure}[H]
  \centering
  \includegraphics[width=0.7\linewidth,clip]{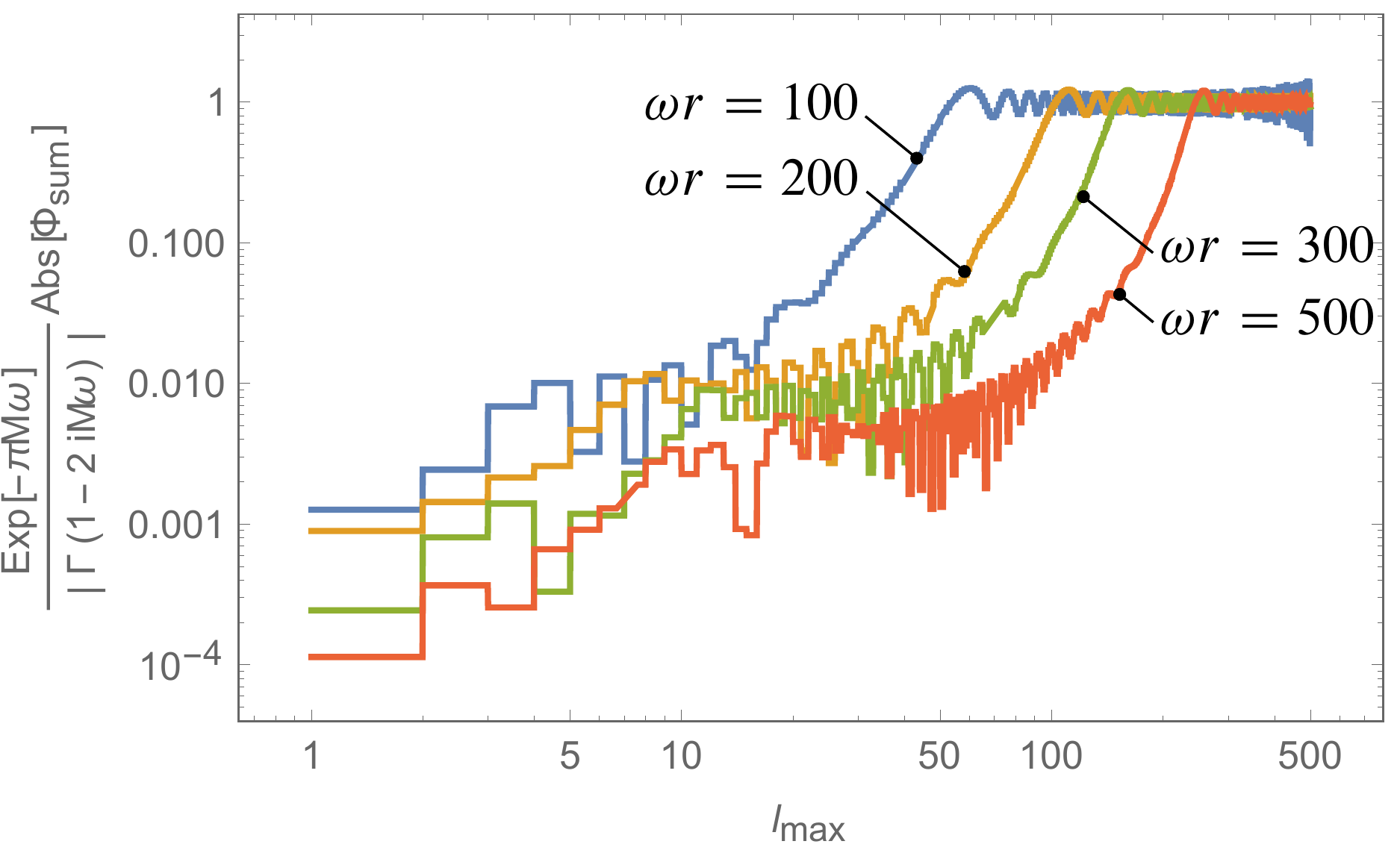}
  \caption{The convergence behavior of the sum (\ref{eq:wave2}) for
    $\theta=0$.}
\end{figure}
%%%
\noindent
We expect the absolute value of this sum becomes $e^{\pi
  M\omega}|\Gamma(1-2iM\omega)|$. Fig.~3 shows the sum converges to
the correct value if we set $\ell_\mathrm{max}=\sqrt{2}\,\omega r$ and
take $\omega r\gg 1$ limit. We also check the sum reproduces correct
$\theta$ dependence of the scattering wave function.
%%%
\begin{figure}[H]
  \centering
  \includegraphics[width=1\linewidth,clip]{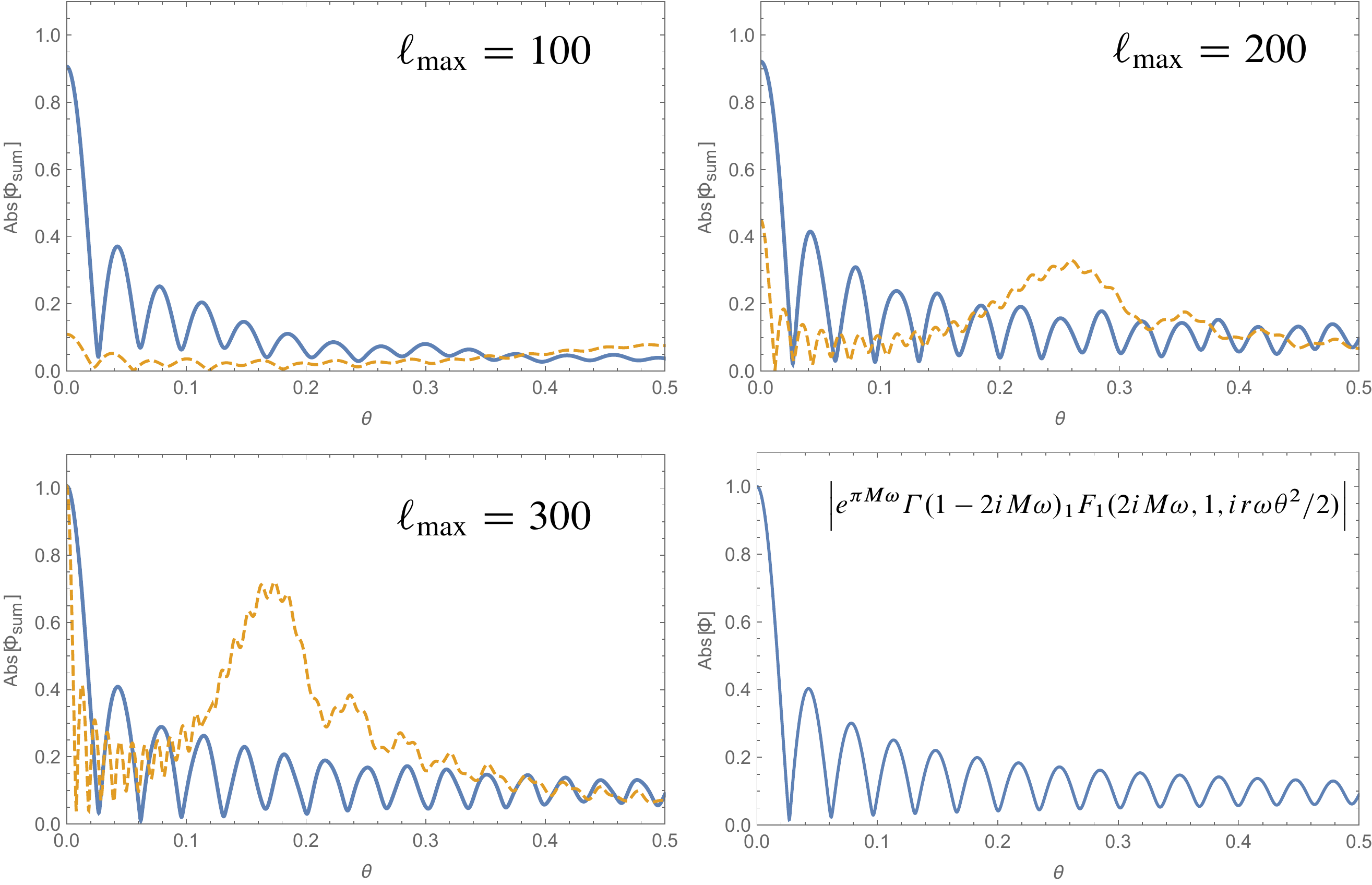}
  \caption{Behaviors of the wave function obtained by the partial wave
    sum up to $\ell_\mathrm{max} (\omega r=200, r=20M)$. The dotted
    lines are the partial wave sum without the factor
    $e^{i\frac{(\ell+1/2)^2}{2\omega r}}$. The last panel is behavior
    of the analytic function that the partial wave sum converges in
    the eikonal limit. }
\end{figure}
%%%
\noindent
Fig.~4 shows behavior of the wave function (\ref{eq:wave2}) obtained
by taking the partial wave sum up to $\ell_\mathrm{max}$  ($\omega r=200,
r=20M$). For $\ell_\mathrm{max}=300\sim\sqrt{2}\,\omega r$, the
obtained wave function has a good agreement with the asymptotic
analytic formula. 

For the Schwarzschild black hole case,  the WKB phase
shift is evaluated by the formula (\ref{eq:phaseshift}).
%%%
\begin{figure}[H]
  \centering
  \includegraphics[width=1\linewidth,clip]{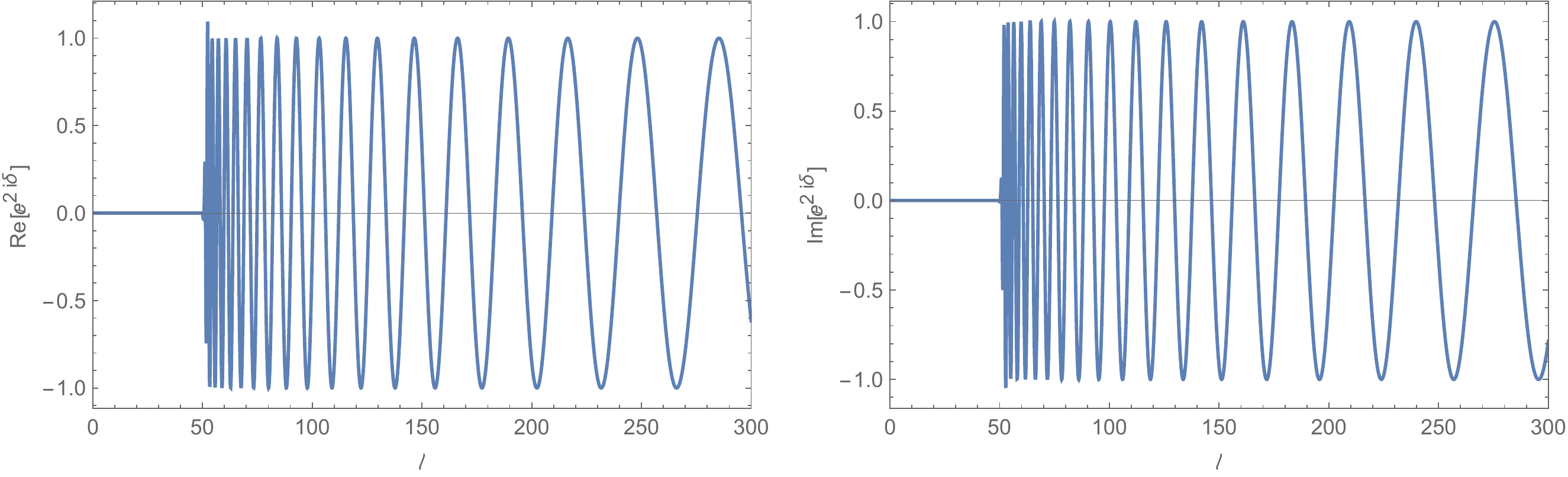}
  \caption{The WKB phase shift of the Schwarzschild black hole for
    $M\omega=10$.}
\end{figure}
%%%
\noindent
Fig.~5 shows appearance of the critical angular number
$\ell_c\approx 51$ corresponding to the critical impact parameter
$b_c=\ell_c/\omega=3\sqrt{3}\,M$ for null rays. To make connection
with the geometric optics, we plot the deflection angle introduced by
the relation (\ref{eq:deflection}):
%%%
\begin{figure}[H]
  \centering
  \includegraphics[width=0.7\linewidth,clip]{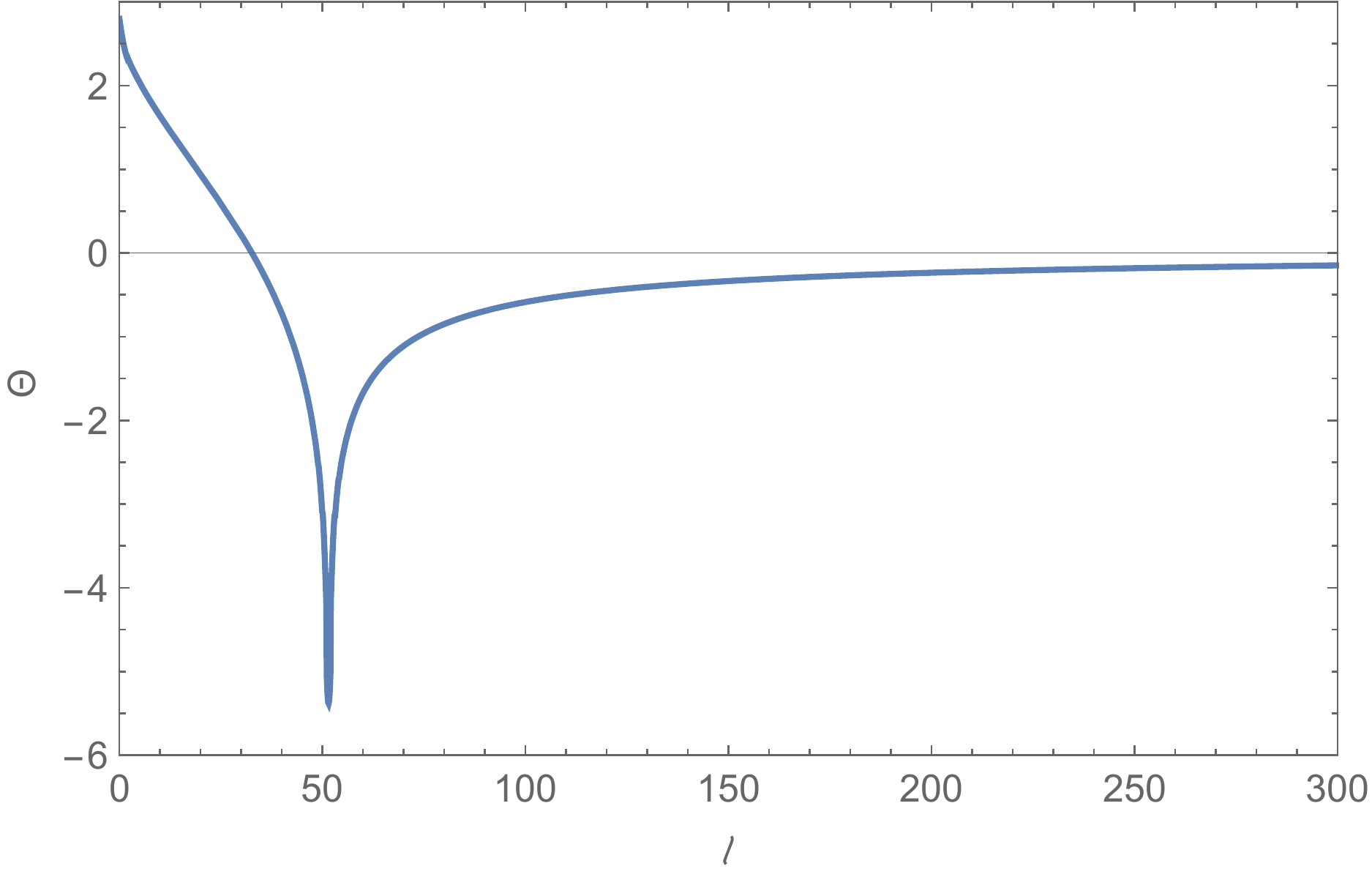}
  \caption{The deflection angle for null rays in the Schwarzschild
    spacetime calculated from the WKB phase shift ($M\omega=10$). The
    critical angular number is $\ell_c\approx 51$.}
\end{figure}
%%%%
\noindent
Near the critical angular number $\ell_c$, the deflection angle can
have arbitrary negative values and  null rays can go around the
black hole arbitrary times (orbiting, Fig.~6). This feature is peculiar to
black hole spacetimes that have event horizons. Fig.~7 shows the wave
function calculated by taking the partial wave sum ($r=20M$).
%%%
\begin{figure}[H]
  \centering
  \includegraphics[width=1\linewidth,clip]{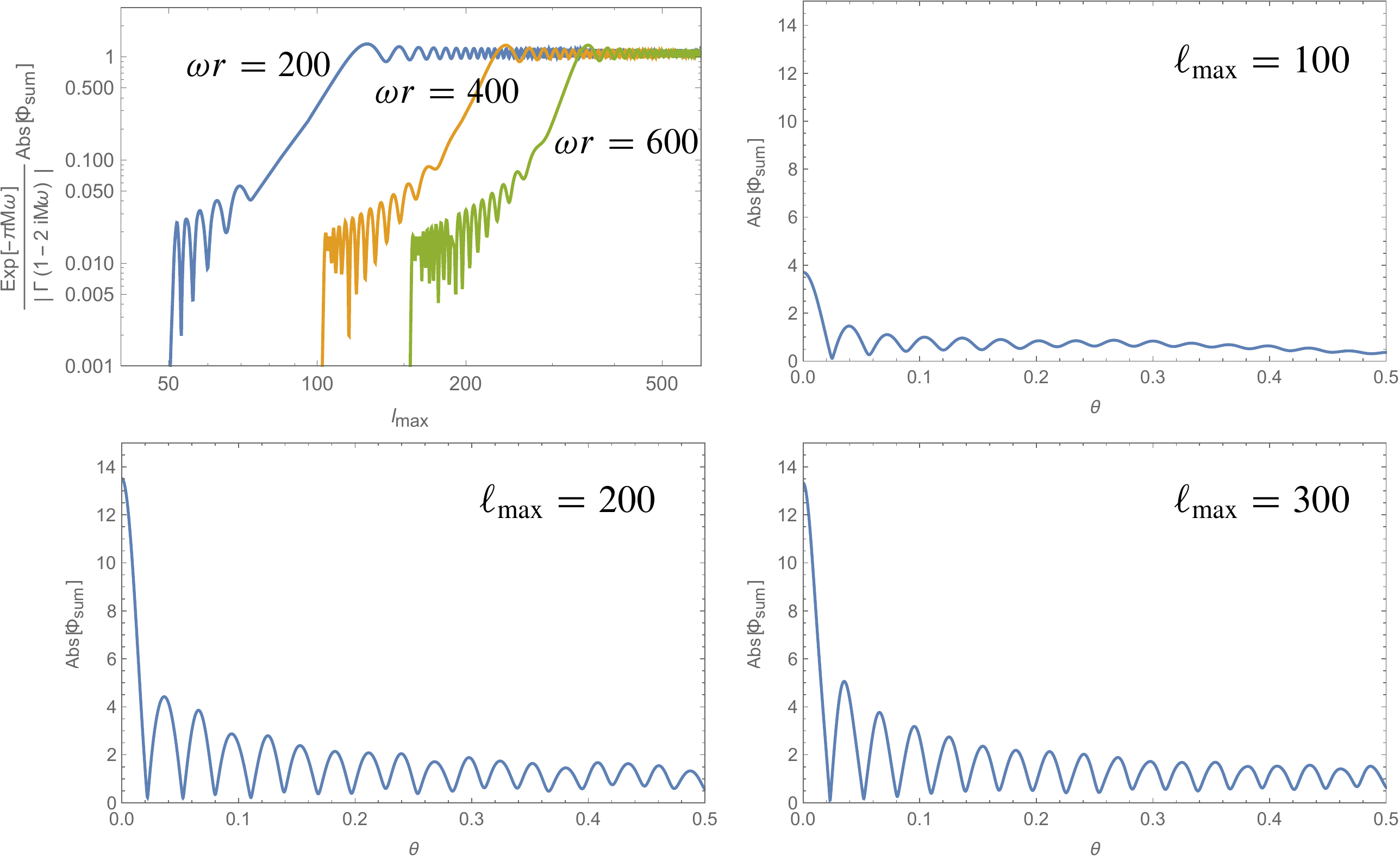}
  \caption{Behavior of the scattering wave function in the
    Schwarzschild spacetime obtained by the partial wave sum. The top
    left panel shows convergence of the wave function for $\theta=0$
    ($r=20M$). The other panels are $\theta$ dependence of the wave
    function with $M\omega=10, r=20M$ and
    $\ell_\mathrm{max}=100,200,300$.}
\end{figure}
%%%
\noindent
We expect the optimal value of $\ell_\mathrm{max}$ for
$M\omega=10, r=20M$ is $\sqrt{2}\,\omega r\approx 280$. Indeed, the
value of the wave function for $\theta=0$ is converging to the expected
value for $\ell_\mathrm{max}>200$ (the top left panel in Fig.~7).

To summarize, base on these examples investigated in this subsection,
we conclude that the partial wave sum formula (\ref{eq:wave1}) is
applicable to black hole spacetimes if we set the upper limit of the
sum as $\ell_\mathrm{max}=\sqrt{2}\, \omega r$ and take the eikonal
limit $\omega r\gg 1$.  In this paper, we try to evaluate this sum
analytically.

%%%%%%%%%%%%%
\subsection{Derivation of lens equation}
Applying the Poisson sum formula again to (\ref{eq:wave1}), the
asymptotic form of the Green function is
%%%%%%%%%%%%%%%%%
\begin{equation}
\fl
\Phi\approx\frac{e^{i\omega(r_*+r_{s*})}}{4\pi i\omega
    rr_s}\sum_{m=-\infty}^{+\infty}
  \int_0^\infty d\lambda\,\lambda\,
  e^{i\frac{\lambda^2}{2\omega
      \tilde r}}\,e^{2i\del_{\lambda-1/2}}\,J_0(\lambda\theta)\,e^{i2\pi
    m(\lambda-1/2)},\quad\mathrm{for}\quad
  \theta\ll1.\label{eq:waveF1}
\end{equation}
%%%
Changing the integration variable to the semi-classical impact
parameter $b=(\ell+1/2)/\omega$, the wave function (\ref{eq:waveF1})
becomes
%%%
\begin{equation}
  \Phi\propto \sum_{m=-\infty}^{+\infty}\int_0^\infty
  db\,b\,e^{i\omega\frac{b^2}{2}\left(\frac{1}{r}+\frac{1}{r_s}\right)}
  \,e^{2i\del_{b\omega-1/2}}\,J_0(b\omega\theta)\,
  e^{i2\pi m(b\omega-1/2)}.
\end{equation}
%%%
As the integrand is the rapidly oscillating function of $b$ for
$M\omega\gg1$, the integral can be evaluated by the stationary phase
method. Using the asymptotic form of the Bessel function\footnote{$$
  J_0(x)\sim \sqrt{\frac{2}{\pi x}}\,\cos(x-\pi/4)\quad\mathrm{for}\quad
  |x|\gg 1.$$}, the stationary phase condition for the integrand yields
%%%
\begin{equation}
  \label{eq:lensEq}
  b\left(\frac{1}{r}+\frac{1}{r_s}\right)
  =-\Theta(b)-2\pi m\pm\theta,\quad
  \Theta(\lambda)\equiv2\frac{d\del_\lambda}{d\lambda}.
\end{equation}
%%%
$\Theta(b)$ is the deflection function which coincides with the
classical deflection angle (\ref{eq:dfangle}).  This is the lens
equation for the black hole spacetime.  For a given position of the
source and the observer $(r,r_s,\theta)$, this equation provides the
impact parameter $b$ and determines the trajectory of null rays
connecting the source and the observer. The integer $m$ has the
meaning of the winding number for null rays orbiting around the black
hole. The geometric interpretation of this equation is shown in
Fig.~8:
%%%
\begin{figure}[H]
  \centering
  \includegraphics[width=0.45\linewidth,clip]{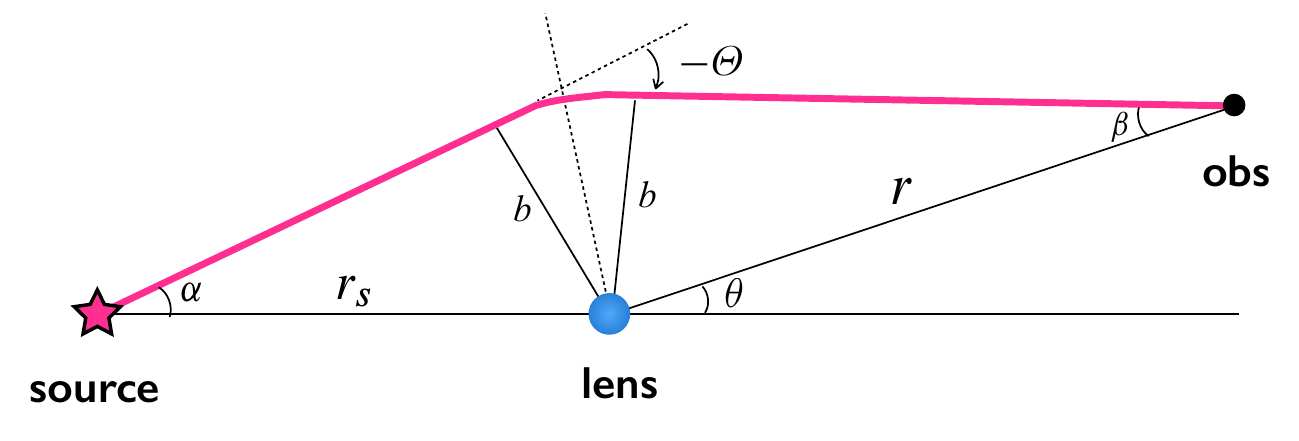}
  \includegraphics[width=0.45\linewidth,clip]{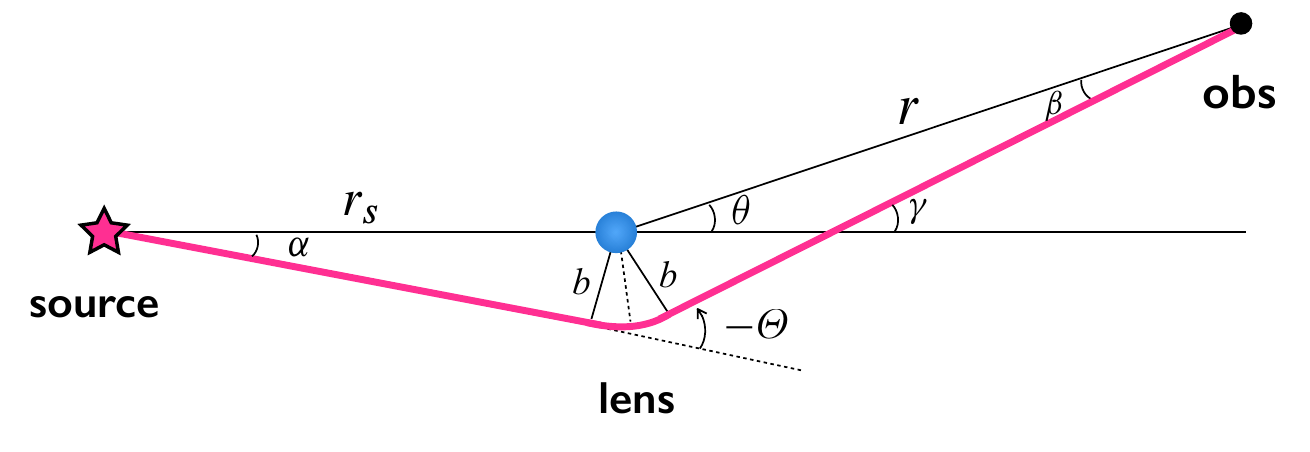}
  \caption{Configuration of the gravitational lensing in the geometric
    optics. Angles in the figure satisfy $\al+\beta=-\Theta+\theta$ in
    the left panel and $\al+\beta=-\Theta-\theta$ in the right panel.
  }
\end{figure}
%%%
\noindent
Angles $\al,\beta$ satisfies the relation
%%%
\begin{equation}
 \al+\beta=-\Theta\pm\theta,
\end{equation}
%%%
where $+$ sign corresponds to the left panel and $-$ sign
corresponds to the right panel in Fig.~8.  Using
$\al\approx\frac{b}{r_s}\ll1$ and $\beta\approx\frac{b}{r}\ll1$, we
obtain
%%%
\begin{equation}
 b\frac{r_s+r}{rr_s}=-\Theta\pm\theta.
\end{equation}
%%%
This is the lens equation (\ref{eq:lensEq}) with $m=0$. For
$-\Theta>2\pi$, the null ray goes around the black hole more than one
times (orbiting). In such a case, the winding number $m$ is defined so
as the condition $0<-\Theta-2\pi m<2\pi$ is satisfied.

%%%%%%%%%%%%%%%%%%%%%%%%%%%%%%%%%%%%
\section{S-matrix and Regge poles for wave scattering by black
  holes}
%%%
For the partial waves with impact parameter $b\sim 3\sqrt{3}M$, the
scattering occurs in the vicinity of the peak  of the effective 
potential (Fig.~9). 
%%%
\begin{figure}[H]
  \centering
  \includegraphics[width=0.5\linewidth,clip]{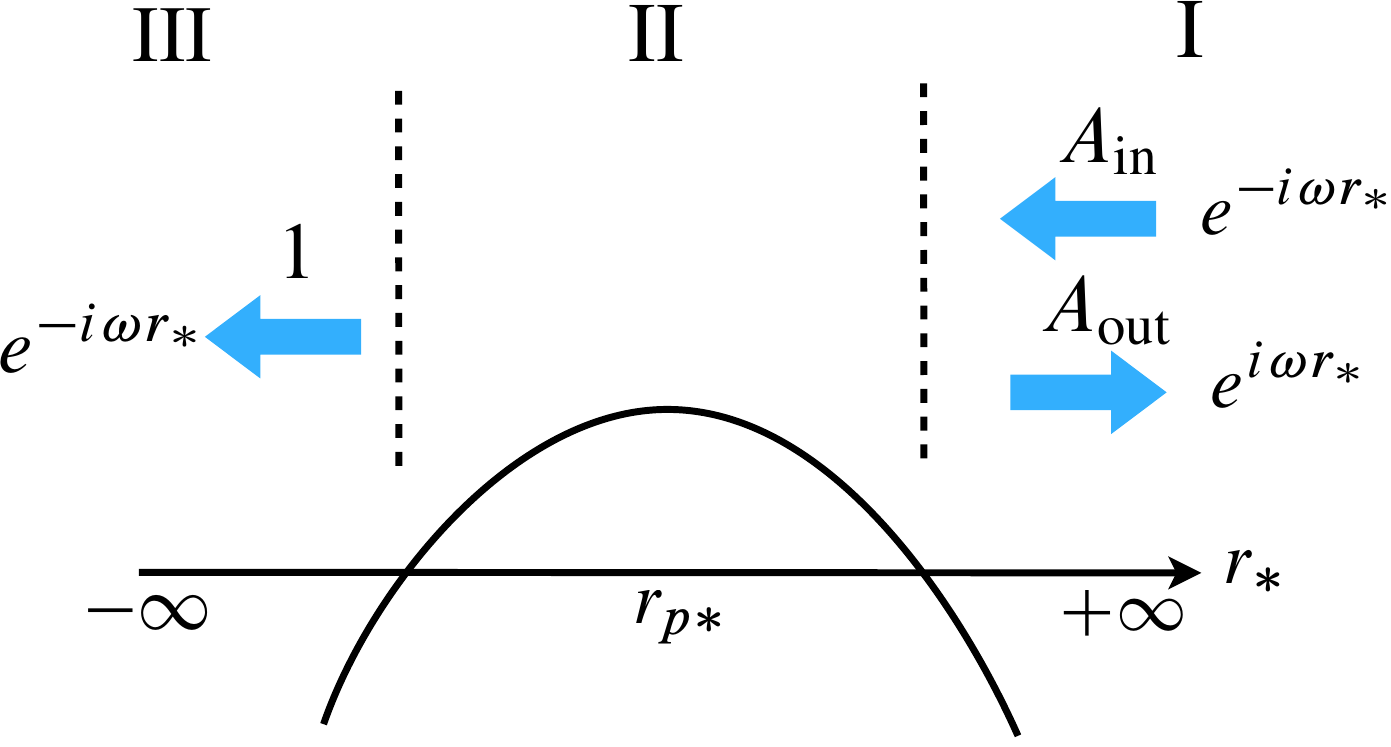}
  \caption{The effective potential $Q(x)$ around the unstable circular orbit
    $r_p$. The quasi-normal mode satisfies
    $A_\mathrm{out}/A_\mathrm{in}=\infty$.}
\end{figure} 
%%%
\noindent
Using the result of the asymptotic matching of the WKB wave function
around the peak, the S-matrix $S_{\ell}\equiv e^{2i\del_\ell}$
is obtained as
%%%
\begin{equation}
 \label{eq:S}
 S_\ell=-(-)^{\ell}\frac{A_{\mathrm{out}}}{A_{\mathrm{in}}}=-(-)^{\ell}
 \frac{e^{i\pi\nu}}{\sqrt{2\pi}}\left(\nu+\frac{1}{2}\right)
^{\nu+1/2}e^{-(\nu+1/2)}\Gamma(-\nu),
\end{equation}
%%%
where $\nu$ is determined by the relation~\cite{Iyer1987}
%%%
\begin{equation}
 \nu+\frac{1}{2}=i\frac{Q_0}{\sqrt{2Q_0''}}\approx
 \frac{27M^2\omega^2-\ell^2}{2\,
   \ell}.
\end{equation}
%%%
The peak of the potential for the Schwarzschild case is at $r_p=3M$ and
%%%
\begin{equation}
 Q_0=Q(r_p),\quad Q''_0=\left.\pa^2_{r_*}Q\right|_{r=r_p}.
\end{equation}
%%%
Absolute value of $S_\ell$ represents the reflection rate of incident waves. 
As the black hole absorbs waves, the phase shift acquires the
imaginary part and the reflection rate becomes smaller than unity for small
impact parameters. Indeed,
%%%
\begin{equation}
 |S_\ell|^2=e^{-2\del_I}=\frac{1}{1+\exp\left
     (-\pi\frac{\ell^2-27(M\omega)^2}{2\ell}\right)}\approx
 \frac{1}{1+\exp\left(-\pi(\ell-\ell_c)\right)},
\end{equation}
%%%
where $\ell_c=3\sqrt{3}M\omega$ is the critical angular momentum.  The
waves with $\ell<\ell_c$ are absorbed by the black
hole. $b_c=\ell_c/\omega=3\sqrt{3}M$ is the critical impact parameter
(Fig.~10).
%%%
\begin{figure}[H]
  \centering
  \includegraphics[width=0.45\linewidth,clip]{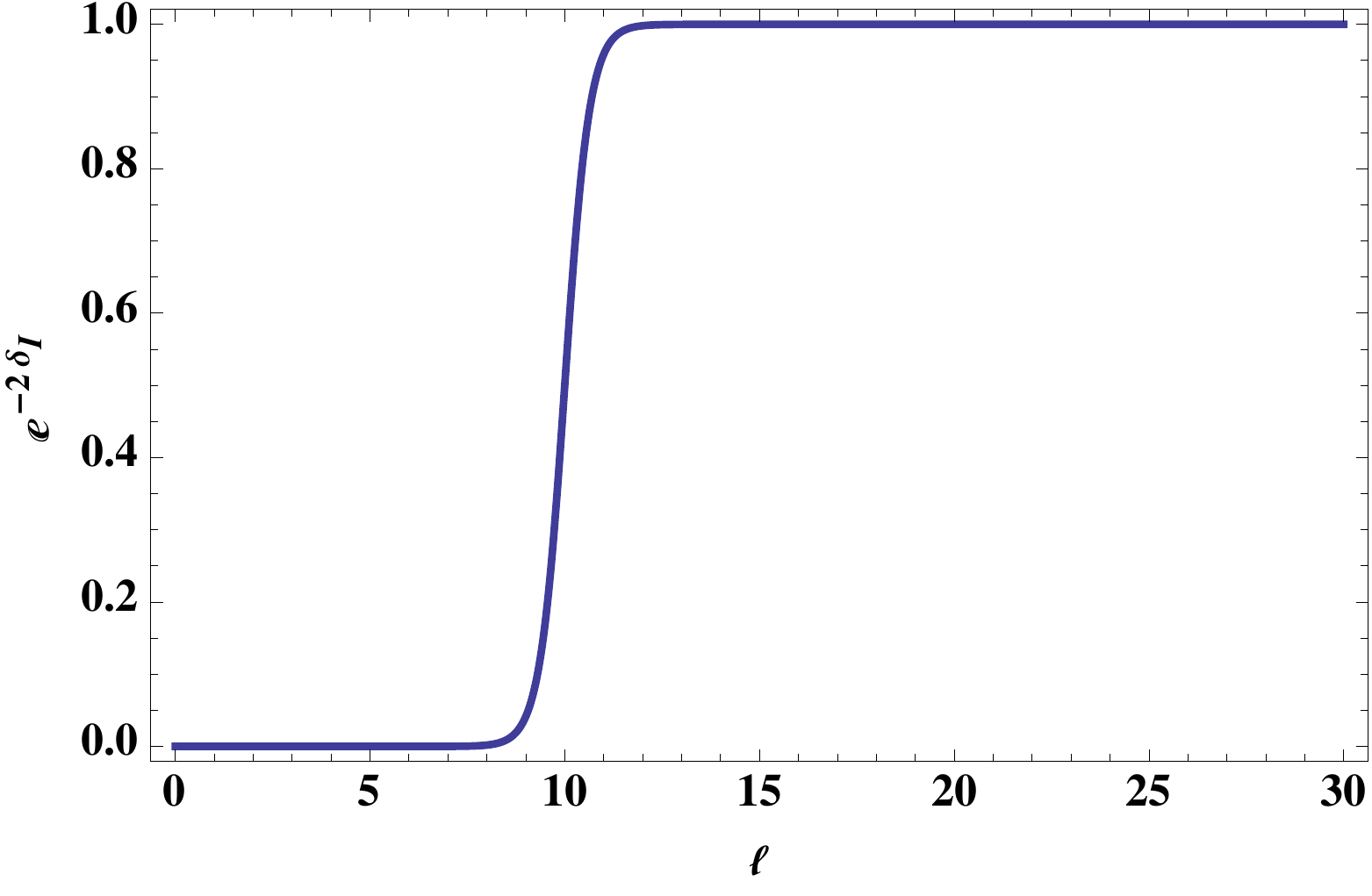}
  \caption{The reflection rate for $\ell_c=10$.}
\end{figure}
%%%
The S-matrix (\ref{eq:S}) has poles in the complex $\ell$ plane of
which location is determined by poles of the Gamma function
$\Gamma(-\nu), \nu=n,~n=0,1,2,\cdots$. That is, poles of the S-matrix
in the complex $\ell$ plane are \footnote{The quasi-normal frequency
  is given by treating $\ell$ as real:
$$
 \omega_n=\frac{1}{3\sqrt{3}M}-\frac{i}{3\sqrt{3}M}\left(n+\frac{1}{2}\right).
$$
}
%%%
\begin{equation}
 \ell_n\approx\ell_c+i\left(n+\frac{1}{2}\right),\quad\ell_c=3\sqrt{3}M\omega,
\quad n=0,1,2,\cdots
\end{equation}
%%%
These are Regge poles~\cite{Newton1982}. In the vicinity of the $n$th pole $\ell_n$, the
S-matrix is
%%%
\begin{equation}
 S_\ell\approx\frac{\gamma_n}{\ell-\ell_n},\quad\gamma_n=
 (-)^{\ell_c}
 \frac{i}{\sqrt{2\pi}}\left(\frac{\ell_n}{\ell_c}\right)\left(n+\frac{1}{2}
\right)^{n+1/2}e^{-(n+1/2)}\frac{1}{n!}.
\end{equation}
%%%
For $n\gg 1$, $n!\approx\sqrt{2\pi n}\,e^{-n}n^n$,
$(1+1/(2n))^{n+1/2}\approx e^{1/2}$, hence the residue $\gamma_n$ is
%%%
\begin{equation}
 \gamma_n\approx
 (-)^{\ell_c}\frac{i}{2\pi}\frac{\ell_n}{\ell_c}.
\end{equation}
%%%%%%%%%%%%%%%%%%%%%%%%%%%%%%%%%%%%%%%%%%%
\section{Evaluation of scattering wave}

The formula of the Green function (\ref{eq:waveF1}) is
decomposed to $m=0$ and $m\neq 0$ parts:
%%%%%%%%%%%%%%%%%
\begin{equation}
\fl
\Phi=\frac{e^{i\omega(r_*+r_{s*})}}{4\pi i\omega
    rr_s}\left[
    \int_0^\infty d\lambda\,\lambda\,e^{i\frac{\lambda^2}{2\omega
        \tilde r}}\,e^{2i\del_{\lambda-1/2}}\,J_0(\lambda\theta)+
    \sum_{m\neq 0}\int_0^\infty
    d\lambda\,\lambda\,e^{i\frac{\lambda^2}{2\omega
        \tilde r}}
    \,e^{2i\del_{\lambda-1/2}}\,J_0(\lambda\theta)\,e^{i2\pi m(\lambda-1/2)}
    \right]. \label{eq:waveF2}
\end{equation}
%%%
As we have discussed in Sec.~2.3, the integer $m$ is the winding
number for null rays orbiting around the black hole. Hence the first
integral term represents contribution of rays which directly reach the
observer (direct part).  By identifying the phase shift as the lens
potential of the gravitational lensing, the direct part reproduces the
Kirchoff-Fresnel diffraction formula for the gravitational
lensing~\cite{Schneider1992}.  On the other hand, the second integral
term represents the contribution of rays go around the black hole
(winding part). We evaluate these two parts separately.
%%%%%%%%%%%%%%%%%%%%%%%%%%%%%%
\subsection{Winding part : contribution of Regge poles}
As the S-matrix has poles in the complex $\ell$ plane, the integral of
the second term in (\ref{eq:waveF2}) can be evaluated by applying
Cauchy's theorem. We use the contour shown in Fig.~11.  We first show
that the integral along the circle $|\lambda|=R$ vanishes as
$R\rightarrow\infty$.  Introducing a new integration variable
$y^2=-i\lambda^2/(2\omega \tilde r)$,\footnote{To show that the
  integral along the circle vanishes, we use the formula
$$
e^{-y^2}=\frac{1}{\sqrt{\pi}}\int_{-\infty}^{+\infty}dk\,e^{-k^2+2iky}.$$}
%%%
%%%
\begin{eqnarray*}
  &\int_{|\lambda|=R} d\lambda\lambda  e^{i\frac{\lambda^2}{2\omega
      \tilde r}}e^{2i\del_{\lambda-1/2}}J_0(\lambda\theta)e^{i2\pi m(\lambda-1/2)}
  \\
&=
  \frac{1}{\sqrt{\pi}}\int_{-\infty}^{+\infty}dk e^{-k^2}\int_{|\lambda|=R}
  d\lambda \lambda\exp\left[\left(\frac{2k e^{i\al}}{\sqrt{2\omega
          \tilde r}}+i2\pi m\right)\lambda
  \right]e^{2i\del_{\lambda-1/2}}J_0(\lambda\theta)e^{-i\pi m},
\end{eqnarray*}
%%%
where we choose $\al=5\pi/4$ for $\mathrm{Im}(\lambda)>0$ and
$\al=-\pi/4$ for $\mathrm{Im}(\lambda)<0$ for convergence of the
integral. Hence, as $R\rightarrow\infty$, the integral becomes zero.
%%%
\begin{figure}[H]
  \centering
  \includegraphics[width=0.35\linewidth,clip]{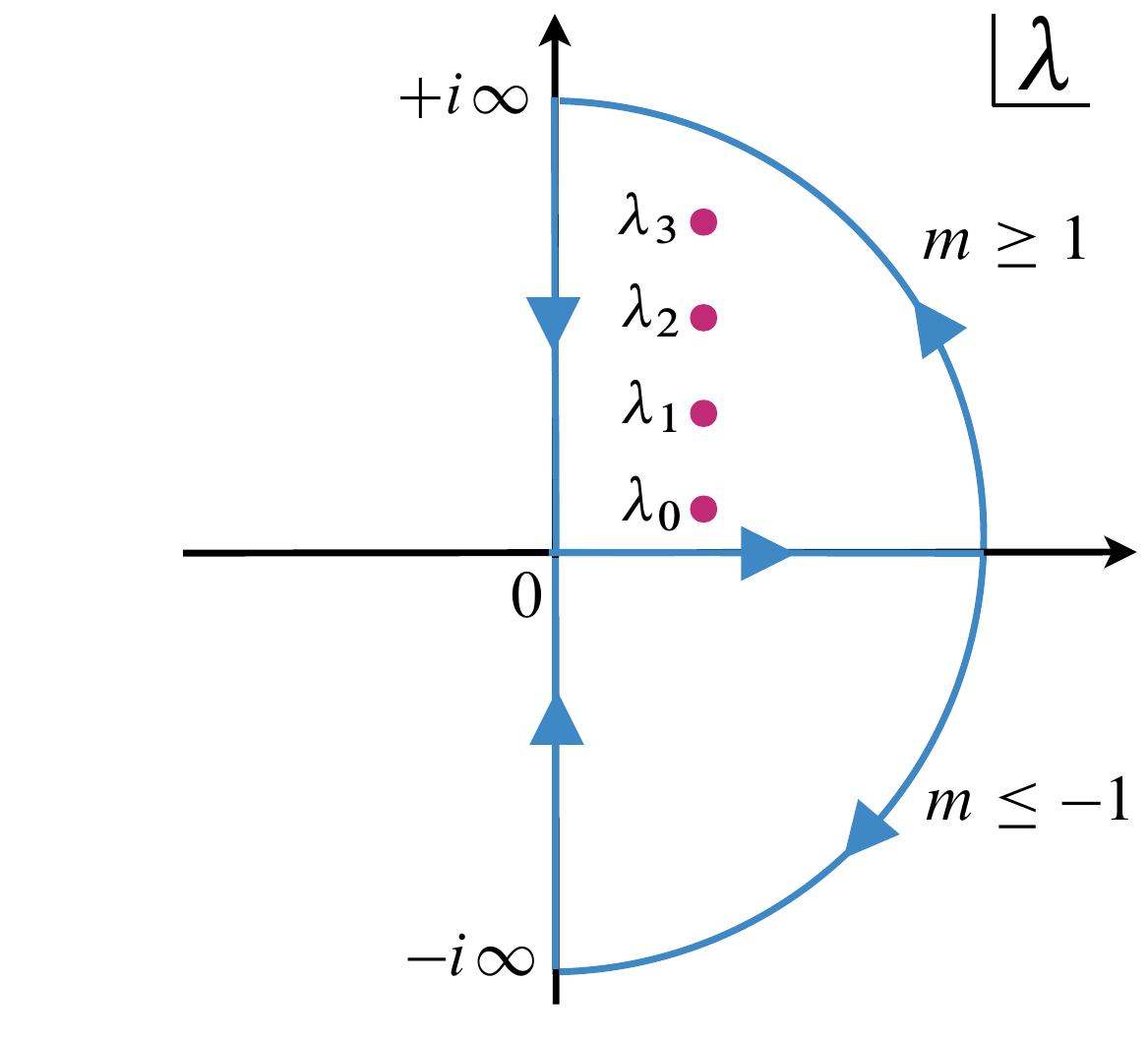}
  \caption{Poles of the S-matrix are situated in the first quadrant of
    the complex $\lambda$ plane.}
\end{figure}
%%%
\noindent
Thus the second integral in (\ref{eq:waveF2}) has contributions
from poles and the integral along the imaginary axis.
%%%
\begin{equation}
 \sum_{m=1}^\infty\sum_{n=0}^\infty\ell_n\,e^{i\frac{\ell_n^2}{2\omega
     \tilde r}}\,(2\pi i\gamma_n)\,J_0(\ell_n\theta)\,e^{i2\pi
   m(\ell_n-1/2)}
 +\sum_{m=1}^\infty\int_0^{+i\infty}d\lambda+\sum_{m=-1}^{-\infty}
\int_0^{-i\infty}d\lambda.
\end{equation}
%%%
The integral along the positive imaginary axis is
%%%
\begin{eqnarray}
  \sum_{m=1}^\infty\int_0^{+i\infty}&\approx
  \int_0^{+i\infty}d\lambda\lambda\frac{i}{2}
  \frac{e^{i\pi(\lambda-1/2)}}{\sin(\pi(\lambda-1/2))}\,
  e^{i\frac{\lambda^2}{2\omega \tilde
                                      r}}\,e^{2i\del_{\lambda-1/2}}J_0(\lambda\theta)
  \nonumber \\
  &=-\frac{i}{2}\int_0^\infty
  d\tilde\lambda\,\tilde\lambda\,\frac{(-i)e^{-\pi\tilde\lambda}}
  {1/2(e^{-\pi\tilde\lambda}+e^{\pi\tilde\lambda})}\,e^{-i\frac{\tilde\lambda^2}
{2\omega \tilde
    r}}\,e^{2i\del_{-i\tilde\lambda-1/2}}J_0(i\tilde\lambda\theta)
 \nonumber \\
 &=O(\omega^0),
\end{eqnarray}
%%%
and yields higher order contribution in the eikonal limit because the
first integral in (\ref{eq:waveF2}) gives $O(\omega)$
contribution. The contribution of the integral along the negative
imaginary axis is also $O(\omega^0)$. We ignore these terms in our
treatment. Thus, the wave function (\ref{eq:waveF2}) becomes
%%%
\begin{equation}
\Phi\approx\frac{e^{i\omega(r_*+r_{s*})}}{4\pi i\omega rr_s}
\left[\int_0^\infty d\lambda\,\lambda\,e^{i\frac{\lambda^2}{2\omega
      \tilde r}}\,e^{2i\del_{\lambda-1/2}}\,J_0(\lambda\theta)+2\pi i
  \sum_{n=0}^\infty\ell_n\,\gamma_n\,e^{i\frac{\ell_n^2}{2\omega
      \tilde r}}\,J_0(\ell_n\theta)\,f(\ell_n)\right],
\end{equation}
%%%
where 
%%%
\begin{equation}
 f(\ell_n)\equiv\sum_{m=1}^\infty e^{2i\pi m(\ell_n-1/2)}
=-\frac{e^{2i\pi\ell_n}}
 {e^{2i\pi\ell_n}+1}.
\end{equation}
%%%
The sum over $n$ can be evaluated by applying the
Euler-Maclaurin formula  and keeping the contribution of the leading order  in
the eikonal limit:
%%%
\begin{eqnarray}
\fl   2i\pi
   \sum_{n=0}^\infty\ell_n\,\gamma_n\,e^{i\frac{\ell_n^2}{2\omega
       \tilde r}}\,J_0(\ell_n\theta)\,f(\ell_n) \nonumber \\
\fl   \qquad\approx
  - \frac{e^{i\pi\ell_c}}{\ell_c}\int_0^\infty
   dn\left(\ell_c+\frac{i}{2}+in\right)^2e^{\frac{i}{2\omega\tilde{r}}\left(
       \ell_c+\frac{i}{2}+in\right)^2}J_0\left[\left(\ell_c+\frac{i}{2}
       +in\right)\theta\right]f\left(\ell_c+\frac{i}{2}+in\right) \nonumber\\
\fl   \qquad\approx 
   \sqrt{\frac{\pi}{2}}\,e^{-\pi-i\pi/4+i3\pi\ell_c}
   e^{\frac{i\ell_c^2}{2\omega\tilde{r}}}\,\ell_c\sqrt{\omega\tilde{r}}\,
   J_0(\ell_c\theta)+O\left(\frac{1}{\sqrt{\omega\tilde r}}\right).
\end{eqnarray}
%%%
We have used
%%%
\begin{equation}
 f\left(\ell_c+\frac{i}{2}\right)\approx -e^{2i\pi\ell_c}e^{-\pi}.
\end{equation}
%%%
After all, we obtain
%%%
\begin{equation}
  \label{eq:waveF3}
 \fl \Phi\approx\frac{e^{i\omega(r_*+r_{s*})}}{4\pi i\omega
    rr_s}
  \left[\int_0^\infty d\lambda\,\lambda\,
    e^{i\frac{\lambda^2}{2\omega\tilde r}}\,e^{2i\del_{\lambda}}
    J_0(\lambda\theta)
+\sqrt{\frac{\pi}{2}}\,e^{-\pi-i\pi/4+i3\pi\ell_c}
e^{\frac{i\ell^2_c}{2\omega\tilde{r}}}\,\ell_c
  \sqrt{\omega\tilde{r}}
   J_0(\ell_c\theta)
    \right].
\end{equation}
%%%

%%%
%%%%%%%%%%%%%%%%%%%%%%%%%%%%%%%
\subsection{Direct part and total wave function}
We evaluate this term assuming that impact parameters of corresponding
null rays are not so small and gravitational field can be approximated
by the Newtonian weak field. Within such an assumption, the phase
shift can be approximated by that of the weak field form
(\ref{eq:Nphase}) and the direct part can be obtained as
(\ref{eq:wave2}). Therefore, the formula for the total scattering wave
is summarized as follows:
%%%
\begin{eqnarray}
\fl  \Phi\approx\frac{e^{i\omega(r_*+r_{s*})}}{4\pi(r+r_s)}\Bigl[
      c_1J_0(b_E\omega\theta)+c_2J_0(b_c\omega\theta)\Bigr],\quad
      \mathrm{for}\quad\theta\ll1,
 \label{eq:waveF4}\\
\fl c_1=e^{-i2M\omega\ln 2M\omega}e^{\pi
    M\omega}\Gamma(1-i2M\omega)e^{-\frac{i}{4}
    \omega\tilde r\theta^2},\quad
   c_2=-i\sqrt{\frac{\pi}{2\omega\tilde r}}\,\ell_c\,e^{-\pi}
       \,e^{i\left(\frac{\ell_c^2}{2\omega\tilde
         r}+3\pi\ell_c-\frac{\pi}{4}\right)}, \nonumber
\end{eqnarray}
%%%
where $b_E=\sqrt{4M\tilde r}$ is the Einstein radius and
$b_c=3\sqrt{3}M$ is the critical impact parameter for null rays. This
formula is applicable for $r,r_s\gg 2M$ and $M\omega\gg 1$. Let us
assume that the point source is near the black hole and the observer
is sufficiently far from the black hole ($r\gg r_s$). Then $\tilde
r\approx r_s$.  Of course the location of the source must be far from
the black hole to apply (\ref{eq:waveF4}), we dare to consider such a
situation to investigate the wave optical image and the interference
effect by the black hole. As the impact parameters for the direct rays 
must be larger than those of the winding rays, we must require $b_E>b_c$
which constrains the distance of the point source from the black hole
in our approximation:
%%%
\begin{equation}
 r_s>\frac{27}{4}M\equiv r_c\sim 6.8M.
\end{equation}
%%%
The ratio of two coefficients is
%%%
\begin{equation}
 \frac{|c_2|}{|c_1|}=\frac{3}{2}\sqrt{\frac{3M}{r_s}}\times
 e^{-\pi}=e^{-\pi}\left(\frac{r_s}{r_c}\right)^{-1/2}\approx 0.043\times
 \left(\frac{r_s}{6.8M}\right)^{-1/2}.
\end{equation}
%%%
In the geometrical optics, it is known that the intensity of the winding
ray reduces to the factor $e^{-2\pi}\approx 0.0019$ compared to the
direct ray~\cite{Fro}.
In terms of the amplitude of the wave, this value corresponds to
$e^{-\pi}\approx 0.043$ and is consistent  with the obtained value
$|c_2|/|c_1|$ in the eikonal limit of the scattering wave function.

%%%%%%%%%%%%%%%%%%%%%%%%%%%%%%%% 
\section{Applications}
As  applications of the analytic formula (\ref{eq:waveF4}) of the
scattering wave, let us consider images of the
black hole and the interference effect in the frequency domain.

%%%%%%%%%%%%%%%%%%%%%%%%%%%
\subsection{Wave optical image of black holes}
Using the scattering wave function (\ref{eq:waveF4}), we can
reconstruct wave optical images of black holes illuminated by a point
source~\cite{Kanai2013}. As the imaging system, we consider a detector
with a convex lens which transforms the interference fringe to the
image (Fig.~12).
%%%%
\begin{figure}[H]
    \centering
    \includegraphics[width=0.4\linewidth,clip]{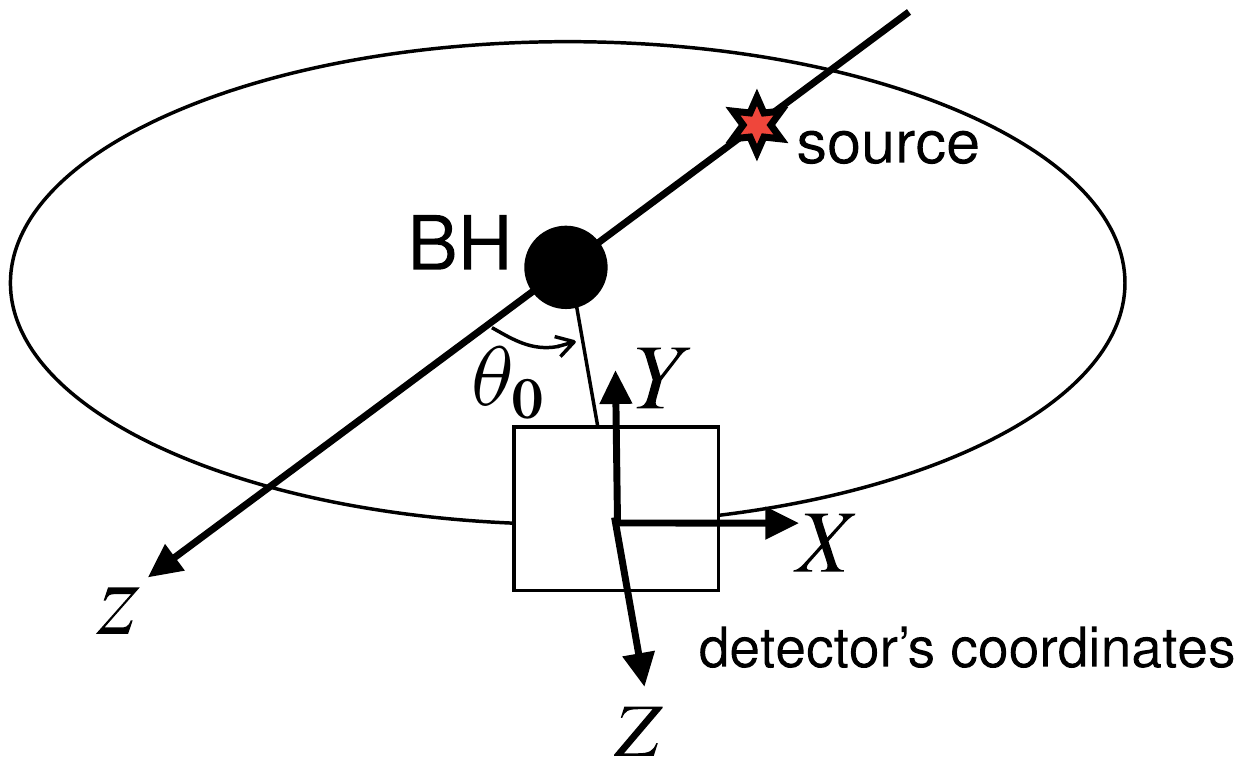}
    \caption{The configuration of the imaging system with a
      convex lens.}
\end{figure}
%%%
\noindent
The location of the lens of the detector is $Z=0$ in the detector's
coordinates and the wave function just in front of the lens is
$\Phi(X,Y,0)$. Using the Fresnel-Kirchoff diffraction
formula~\cite{Sharma2006}, the wave on the detector's focal plane
$Z=f$ is given by
%%%
\begin{equation}
    \label{eq:lens1}
    \Phi_I(X_I,Y_I)\propto\int_{X^2+Y^2\le
      a^2}dXdY\,\Phi(X,Y,0)\,
 e^{-i\omega\frac{X^2+Y^2}{2f}}\times\frac{e^{i\omega\ell}
    }{\ell}
\end{equation}
%%%
where $X_I,Y_I$ are coordinates on the focal plane, $a$ is a radius of
the lens and $\ell$ is the path length between a point on the lens
plane and a point on the focal plane.  The function
$e^{-i\omega\frac{X^2+Y^2}{2f}}$ in (\ref{eq:lens1}) represents the
action of the convex lens which transform incident plane waves to
spherical waves focusing at the focal point.  Assuming that the focal
length is sufficiently larger than the radius of the lens,
%%%
\begin{equation}
 \ell=\sqrt{(X-X_I)^2+(Y-Y_I)^2+f^2}\approx f+\frac{(X-X_I)^2+(Y-Y_I)^2}{2f},
\end{equation}
%%%
and we obtain
%%%
\begin{equation}
    \Phi_I(X_I,Y_I)\propto\int_{X^2+Y^2\le a^2}dX dY\,\Phi(X,Y,0)\,
    e^{-i\frac{\omega}{f}(X_IX+Y_IY)}.
\end{equation}
%%%
Thus the wave function $\Phi_I$ on the focal plane is the Fourier
transformation of the wave function $\Phi$. The relation between the
world coordinates $(x,y,z)$ and the detector's coordinates $(X,Y,Z)$
is
%%%
\begin{eqnarray}
    &x=r\sin\theta\cos\phi=X\cos\theta_0-Z\sin\theta_0, \nonumber \\
    &y=r\sin\theta\sin\phi=Y,\\
    &z=r\cos\theta=X\sin\theta_0+Z\cos\theta_0,\quad X^2+Y^2+Z^2=r^2. \nonumber
\end{eqnarray}
%%%
It is possible to obtain $\cos\theta$ as a function of $X,Y$:
%%%
\begin{equation}
    \cos\theta=\left(\frac{X}{r}\right)\sin\theta_0
    +\sqrt{1-\left(\frac{X}{r}\right)^2-\left(\frac{Y}{r}\right)^2}\,
    \cos\theta_0.
\end{equation}
%%%

Fig.~13 shows an example of images of the Schwarzschild black hole
illuminated by a point source. Images are obtained by drawing
$|\Phi_I|$ which is the amplitude of the Fourier transformation of
$\Phi$.
%%%
\begin{figure}[H]
  \centering
   \includegraphics[width=0.9\linewidth,clip]{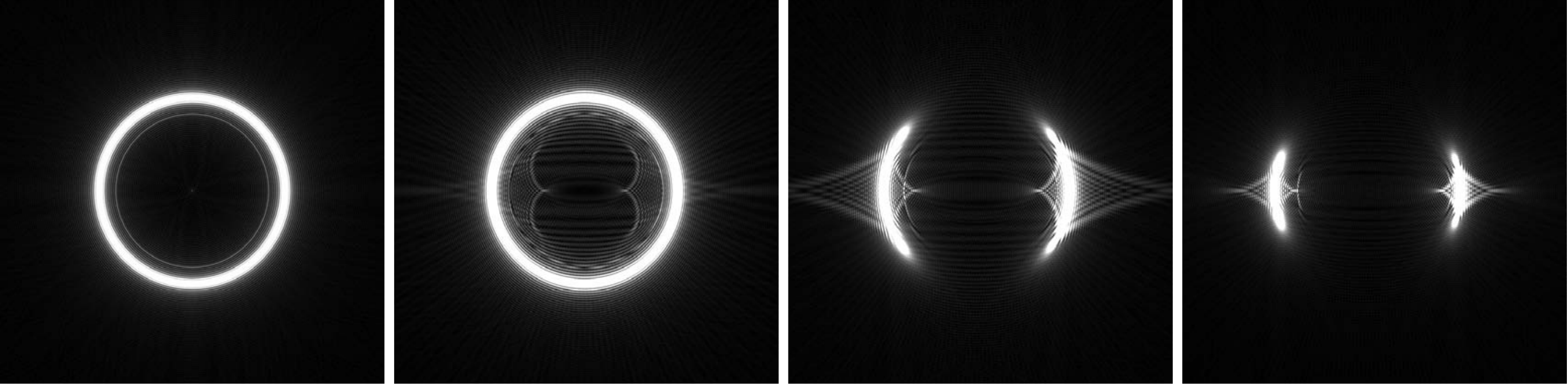}
   \caption{Images of the black hole for different viewing angles
    $\theta_0=0.0, 0.02, 0.05, 0.09$ (from the left to the right
    panels). Parameters are $r_s=10M, r=\infty$. $M\omega=800$ and
    $a=0.03$.}
\end{figure}
%%%
\noindent
In our previous paper~\cite{Kanai2013}, we numerically solved the
Helmholtz equation in the Schwarzschild spacetime and images of the
black hole are reconstructed using the obtained wave function. Due to
limited resolution of the numerical calculation, it was difficult to
perform simulation with high frequency waves corresponding to the
eikonal limit and it was hard to resolve the fine structure of black
hole images. In the present analysis, as we have obtained the analytic
form of the scattering waves, it is possible to acquire images using
high frequency waves. In Fig.~13, we can observe a bright outer ring
which corresponds to the Einstein ring by direct rays and a thin inner
ring which corresponds to the rim of the black hole shadow (photon
ring). It was not easy to recognize this structure of the black hole
images in our previous paper~\cite{Kanai2013}.

%%%%%%%%%%%%%%%%%%%%%%%%%%%%%%%%%%%%%%%%%%%%%%
\subsection{Interference effect in wave scattering by black holes}
We can examine the interference effect in the frequency domain using
(\ref{eq:waveF4}); that is
caused by the interference between the direct rays and the winding
rays.  For $b_{E}\omega\theta,b_c\omega\theta\gg 1$, using the
asymptotic form of the Bessel function, the scattering wave
(\ref{eq:waveF4}) can be written as
%%%
\begin{equation}
 \Phi\sim
 d_1\cos(b_E\omega\theta_0)+d_2\cos(b_c\omega\theta_0),
\end{equation}
%%%
where $d_1,d_2$ represent numerical factors independent of $\omega$
with their ratio
%%%
\begin{equation}
 \frac{|d_2|}{|d_1|}=e^{-\pi}\left(\frac{r_s}{r_c}\right)^{-1/2}.
\end{equation}
%%%
After averaging  the amplitude of the wave over rapidly oscillating
frequency scale $\sim 1/(M\theta_0)$, the power spectrum  is
%%%
\begin{equation}
  I(\omega)=|\Phi|^2\sim\frac{|d_1|^2}{2}+\frac{|d_2|^2}{2}+\frac{d_1d_2^*+
    d_1^*d_2}{2}\cos\left[(b_E-b_c)\omega\theta_0\right].
\end{equation}
%%%
The first term and the second term represent the intensity of the
direct wave and the winding wave, respectively. The third term
represents the interference between the direct wave and the winding
wave.  For the point source at $r_s\sim 7M$, the impact parameters for
the direct rays and the winding rays become nearly equal and the
interference term results in the ``beat'' in the frequency domain. The
modulation due to the interference appears in the power spectrum
(Fig.~14).
%%%%%%%%%%%%%%%%
\begin{figure}[H]
  \centering
  \includegraphics[width=1.04\linewidth,clip]{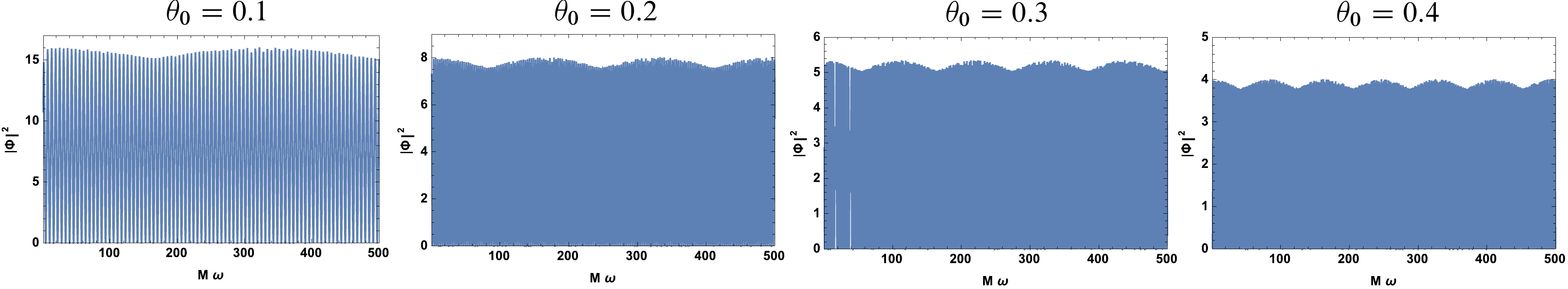}
  \caption{The power spectrum of the scattering wave with
    $r_s=7M$. The amplitude of the modulation is $\Delta I/I\sim 0.1$. The
    period of the modulation depends on the viewing angle $\theta_0$.}
\end{figure} 
%%%
\noindent
The period of the modulation in the power spectrum is
%%%
\begin{equation}
 M\Delta\omega=\frac{2\pi}{\theta_0}
   \frac{1}{3\sqrt{3}}\left(\sqrt{\frac{r_s}{27M/4}}-1\right)^{-1}.
\end{equation}
%%%
As the typical value for the viewing angle $\theta_0$, if the Einstein angle
$\theta_E=\sqrt{4M/r_s}$ is assumed, we obtain
%%%
\begin{equation}
  M\Delta\omega=\frac{\pi}{2-3\sqrt{\frac{3M}{r_s}}},\quad r_s>\frac{27M}{4}.
\end{equation}
%%%%
For $r_s\sim r_c$, the visibility of the modulation is 
%%%
\begin{equation}
  V\equiv\frac{I_{\mathrm{max}}-I_{\mathrm{min}}}{I_{\mathrm{max}}+I_{\mathrm{min}}}
  =\frac{d_1d_2^*+d_1^*d_2}{|d_1|^2+|d_2|^2}\sim 2e^{-\pi}\sim 0.08.
\end{equation}
%%%
This value is not so small and we may have the possibility to detect
this interference effect observationally, which means direct
verification of the black hole spacetime.
%%%
\begin{table}[H]
\centering
\begin{tabular}{|c|c|c|c|} \hline
  & galactic core BH  & intermediate mass BH & stellar mass BH \\
  \hline
  mass  & $10^6M_\odot$ & $10^3M_\odot$ & $10M_\odot$ \\ \hline
 $\Delta\omega$ & 40Hz & 40kHz & 4MHz \\ \hline
\end{tabular}
\caption{The typical frequency of the modulation in the power
  spectrum.}
\end{table}
%%%
\noindent
Table I shows the frequencies $\Delta \omega$ for black holes with
different masses. For intermediate mass black holes and stellar mass
black holes, the modulation in the power spectrum is expected to be
detectable using sub-mm radio telescopes with suitable band width and
the frequency resolution.

%%%%%%%%%%%%%%%%%%%%%%%%%%%%%%%%%%%%%%%%%%%%%%
\section{Summary}
We investigated the wave optics in the Schwarzschild spacetime
following the standard treatment of the wave scattering problem and
evaluated the Green function for the stationary monochromatic point
source in the eikonal limit.  To make the partial wave sum of the
scattering wave converge, we retain the next to leading order
contribution in the phase of the partial waves for large $r$.  Effect
of the orbiting scattering is taken into account as the contribution
of Regge poles to the scattering wave. The wave optical images of the
black hole are obtained and the interference effect in the frequency
domain is discussed.

As the straightforward extension of the analysis performed in this
paper, the formulation of the wave optics in the Kerr spacetime is now
on going. We expect that the effect associated with the black hole
spin such as the superradiance adds new features to the scattered
wave.  We will report on this subject in our forthcoming paper.

%%%%%%%%%%%%%%%%%%%%%%%%%%%%%%%%%%%%%%%%%%%%%%%%%%%%
\ack 
The authors thank the anonymous referee who pointed out the importance
of convergence of the partial wave sum.
This work was supported in part by the JSPS KAKENHI Grant Number
15K05073. The authors also thanks all member of ``black hole horizon
project meeting'' in which the preliminary version of this paper was
presented.

%\References
\vspace{1cm}
\noindent
\textbf{References}
\vspace{0.5cm}

%%%%%%%%%%%%%%%%%%%%%%%%%%%%%%%%%%%%%%%%%%%%%%
%\bibliography{My_projects,BH}
%\bibliography{cosmology,my-paper,relativity,quantum,black-hole}

\end{document}